\documentstyle[psfig]{mn}

\newcommand{\hmpc}{$h^{-1}\ Mpc$}

\title[Recovering the Initial Conditions of our Local
Universe]{Recovering the Initial Conditions of our Local Universe from
NOG and PSCz Catalogues.}  \author[Fontanot et al.]{Fabio
Fontanot$^1$, Pierluigi Monaco$^1$, and Stefano Borgani$^1$\\
$^1$Dipartimento di Astronomia, Universit\`a di Trieste, 
via Tiepolo 11, 34131 Trieste, Italy \\
email: fontanot, monaco, borgani@ts.astro.it}

\begin{document}
\maketitle

\begin{abstract}
We apply the ZTRACE algorithm to the optical NOG and infra-red PSCz
galaxy catalogues to reconstruct the pattern of primordial
fluctuations that have generated our local Universe.  We check that
the density fields traced by the two catalogues are well correlated,
and consistent with a linear relation (either in $\delta$ or in
$\log(1+\delta)$) with relative bias (of NOG with respect to PSCz)
$b_{\rm rel}=1.1\pm 0.1$. The relative bias relation is used to fill
the optical zone of avoidance at $|b|<20^\circ$ using the PSCz galaxy
density field.

We perform extensive testing on simulated galaxy catalogues to
optimize the reconstruction.  The quality of the reconstruction is
predicted to be good at large scales, up to a limiting wavenumber
$k_{\rm lim}\simeq 0.4\ h/Mpc$ beyond which all information is lost.
We find that the improvement due to the denser sampling of the optical
catalogue is compensated by the uncertainties connected to the larger
zone of avoidance.

The initial conditions reconstructed from the NOG catalogue are found
(analogously to those from the PSCz) to be consistent with Gaussian
paradigm. We use the reconstructions to produce sets of initial
conditions ready to be used for constrained simulations of our local
Universe. 
\end{abstract}


\section{Introduction}

In the gravitational instability scenario the observed large-scale
structure of the Universe forms by gravitational collapse of small
perturbations imprinted in the early Universe by some mechanism like
inflation.  The perturbation field is believed to be a Gaussian
stochastic process.  In this case, all the cosmological information
present in the perturbation field is given by its power spectrum, or
by its Fourier transform, the two-point correlation function.  On the
other hand, gravitational evolution induces strong non-Gaussianities
in the large-scale density distribution, easily recognizable both in
the 1-point Probability Distribution Function (hereafter PDF) of the
density, which is very similar to log-normal, and in the topology of
large-scale structure, composed by voids, filaments and clusters (see,
e.g., Sahni \& Coles 1996).

The specific pattern of primordial fluctuations that has originated
the structure observed in our Local Universe bears no cosmological
information beyond that of the power spectrum, but a detailed
knowledge of it is important for many purposes.  Firstly, its
reconstruction relies on the assumption of gravitational instability,
but not on the Gaussianity of the initial field; the reconstructed
density field can then in principle be used to test the Gaussianity of
the initial conditions (Nusser, Dekel \& Yahil 1995, Monaco et
al. 2000).  These test have not revealed any non-Gaussian feature.
However, they are not very sensitive to small signal, and the signal
is degenerate with spurious non-Gaussian features induced by
non-linear bias (Monaco et al. 2000). 
 
Secondly, the reconstruction can be used to run constrained
simulations that reproduce our local Universe.  Such simulations are
useful for at least two aims, namely to calibrate distance indicators
for non-homogeneous Malmquist bias (see, e.g., Strauss \& Willick
1995; Kolatt et al., 1996), and to model galaxy formation in an
environment that closely resembles our local Universe (Narayanan et
al., 2001; Klypin et al. 2002; Mathis et al. 2002).  Indeed, the local
Universe is obviously the place where observations are most detailed,
and where objects such as dwarf galaxies or LBS spirals are mostly
observed. Any study of the environmental dependence of galaxy
properties would greatly benefit from this kind of simulation.

The density field at scales of from a few to a few tens of \hmpc\ is
evolving in the quasi-linear regime, in which the perturbations still
retain memory of the initial conditions; however, the inversion of the
gravitational evolution in this regime is not trivial.  This is due to
many reasons: (i) the evolution has a significant degree of
non-linearity; (ii) the presence of relaxed structure at small scales
influences the reconstruction; (iii) the sampling of the density field
by galaxy catalogues is not homogeneous, due to the selection
function, the presence of unobserved regions (like the Zone of
Avoidance, hereafter ZOA) and to possible inhomogeneities of the
catalogues; (iv) galaxies are biased tracers of the total density
field, and the bias relation is poorly known; (v) distance estimates
are unavailable for large complete galaxy catalogues, so that the
density field is observed in the redshift space.

Standard N-body integrators are not suitable to solve the inversion
problem, as any noise would be interpreted as decaying mode and
amplified in the backward integration.  The proposed solvers are based
either on the Zel'dovich (1970) approximation or on the minimum-action
principle (Peebles 1989, 1990; Branchini \& Carlberg, 1994).  The most
recent proposals are PIZA (Croft \& Gatzagnaga 1997), ZTRACE (Monaco
\& Efstathiou 1999), FAM (Nusser \& Branchini 2000), MAK (Frisch
et al. 2000); older proposals are reviewed in Narayanan \& Croft
(1999).  These algorithms are able to invert the gravitational
evolution on scales larger than 3-5 \hmpc \ (Gaussian smoothing),
where the highly-non linear regime starts to prevail.  The inversion
codes are also used to find the density field in the real-space and
the 3D-field of peculiar velocities.
 
The reconstruction has been attempted by many groups. Kolatt et
al. (1996) used the IRAS 1.2 Jy redshift survey (Fisher et al., 1995),
in order to recover the density field in real space, along with the
velocity and the potential fields. The initial conditions were then
reconstructed by applying the Bernoulli-Zel'dovich equation (Nusser \&
Dekel, 1992) to the peculiar velocity field. Then they forced the
1-point PDF of their resulting initial conditions to be Gaussian by
applying a rank- and variance-conserving procedure. Narayanan et
al. (2001) applied the same approach to the analysis of the PSCz
catalogue and tested their results against simulations for various
cosmologies and bias schemes. Mathis et al. (2002) started their
analysis from the IRAS 1.2 Jy redshift survey. They smoothed heavily
the density field and computed the velocity and the potential
fields. Through the application of Bernoulli-Zel'dovich equation they
reconstructed the initial density field. They used this field as a
constraint for the Hoffmann-Ribak algorithm (Hoffmann \& Ribak, 1991),
to obtain a realization of a Gaussian field that resembles the input
one once smoothed on the same scale. Klypin et al. (2002)
reconstructed the density and velocity fields applying the formalism
of Wiener Filtering (see, Zaroubi et al, 1995) and the Hoffmann-Ribak
algorithm to the MARK III catalogue of peculiar velocities (Willick et
al., 1997). Initial conditions were recovered by applying linear
theory to the velocity field.

In this paper we present a new reconstruction of the local Universe
which improves over the previous ones in many respect.  We use the
ZTRACE code (Monaco \& Efstathiou 1999; Monaco et al. 2000) to trace
back in time the density field obtained from galaxy catalogues.  This
code is able to self-consistently invert the gravitational evolution
preserving the statistics of the initial density field, except in the
high peaks which are dominated by highly non-linear dynamics.  The
code has already been applied to the infra-red IRAS PSCz redshift
catalogue (Saunders et al. 2000; Monaco et al. 2000), with the result
of a Gaussian 1-point PDF of the reconstructed linear density field
and no clear sign of distortion due to non-linear bias.

For this reconstruction we use both the PSCz and the optical NOG
catalogue (Giuricin et al. 2000) to sample the galaxy density field.
The advantage of this optical catalogue with respect to the infra-red
one is an increase of density by a factor of 1.3 to 2 and a better
relation of optical light with galaxy mass, which allows sampling of
the early type, dust-poor galaxy population.  The disadvantage is
given by the larger ZOA and by the inhomogeneity of the sources,
although the work of Marinoni et al. (1999) and Marinoni (2000)
has demonstrated the very good degree of completeness of the sample.
The use of two different catalogues allows us to test the robustness of
the reconstruction to the sampling of the density field by different
types of galaxies.

The paper is organized as follows: section 2 describes the two
catalogues used, and compares the galaxy density fields traced by
them; the relative bias of the two catalogues is quantified, and this
relation is used to fill the optical ZOA with the information given by
the PSCz.  Section 3 describes the tests performed on simulated galaxy
catalogues to assess the accuracy of the reconstructed initial
conditions in the Fourier space; in particular, we quantify the limit
$k$-value at which the reconstruction is not reliable any more, and
the corrections that are required to fix the flattening of the
reconstructed peaks and to restore the power lost by smoothing.
Section 4 shows the consistency of the initial conditions
reconstructed from the NOG catalogue with Gaussianity, and the overall
accuracy with which the structure of the local Universe is
reconstructed by a set of constrained simulations. Finally, section 5
gives the conclusions.

\section{The catalogues}

\subsection{PSCz and NOG}

Galaxies are biased tracers of the underlying density field, and this
bias depends on the properties of the galaxy catalogue used.  The PSCz
catalogue (Saunders et al. 2000) meets the most important requirement,
i.e. the homogeneity of selection, in that it is extracted from the
Point Source Catalogue of the IRAS satellite (Joint IRAS Science
Working Group 1988).  In particular, all sources with 60$\mu$ fluxes
$>0.6$ Jy which show galaxy-like FIR colors have been selected and
re-observed in the optical.  The final catalogue contains about 15000
confirmed galaxies over 84 per cent of the sky.  Due to the
homogeneity of the selection and the smallness of the ZOA (limited
roughly to $|b|<8^\circ$ plus two thin stripes not observed by IRAS)
the PSCz catalogue has been fruitfully used for many studies of
large-scale structure (see Saunders et al., 2000 and references
therein).  Another advantage of the FIR selection relies in the
high-luminosity cutoff of the selection function, which is weaker than
the optical, and allows a sparse sampling to much larger distances.
The main disadvantage relies in the poor relationship between FIR
light, mostly emitted by dust heated up by star formation or AGN, and
galaxy mass.  In other words, the FIR light is biased toward mid-type
spirals (Marinoni, private communication), missing almost completely
the early-type population.

Optical light is more tightly connected to galaxy mass, due to the
presence of relations such as Tully-Fischer or the fundamental plane.
On the other hand, optical all-sky catalogues must necessary rely on
compilations of different sets of observations; this is a source of
systematic error that may be difficult to subtract.  Moreover, the ZOA
in the optical is much wider than the IR.  All-sky optical catalogues
have been presented e. g. by Tully (1988) and Santiago et al. (1995).
We use the NOG catalogue (Giuricin et al. 2000), which has been
selected from the LEDA database (see, e. g., Paturel et al., 1997) and
then homogenized by Garcia et al. (1993), Marinoni et al. (1999)
and Marinoni (2000).  It is 95 per cent complete to $B=14$ and 98 per
cent complete in redshift. It is limited to $|b|>20^\circ$ and
$cz<6000 \ km\ s^{-1}$, after which the selection function drops
abruptly\footnote{The cut in redshift is due in fact to the low degree
of completeness of the LEDA database beyond that limit.}. Giuricin et
al. (2000) have extracted galaxy groups from the catalogue using both
a percolative, friends-of-friends algorithm (Huchra \& Geller, 1982)
and a hierarchical one (Materne, 1978).  The catalogue has been used
by Giuricin et al. (2001) for studies on redshift-space two-point
correlation functions of galaxies and groups, by Girardi et al. (2002)
for the analysis of observational mass-to-light ratio of galaxy
systems and by Marinoni, Hudson \& Giuricin (2002) to obtain the
optical luminosity function of virialized systems.

\subsection{The relative bias between PSCz and NOG}

It is important to quantify how the two catalogues map the galaxy
density field, and in particular their relative bias.  This is
important to interpret possible differences in the reconstruction of
the initial conditions.  Moreover, to obtain a full-sky coverage of
the density field, it is important to fill in the optical ZOA with a
galaxy field which resembles the true one as accurately as possible.
The best way to do it is to use the information given by the PSCz
itself, whose ZOA is much smaller.  This requires an assessment of
the relation between the galaxy density fields as described by the two
catalogues.

We can try to anticipate the result.  The rms of galaxy counts in
spheres of 8 $h^{-1}\ Mpc$, usually denoted as $\sigma_8$, has been
measured to be $\sim 0.9$ for optical galaxies (Szalay et al., 2002)
and $\sim 0.8$ for IRAS ones (Seaborne et al. 1999, Sutherland et
al. 1999). So, as long as both kinds of galaxies sample the density
field with a roughly linear bias scheme, we expect to measure a
relative bias parameter $b_{\rm rel} \simeq \sigma_{8,{\rm opt}} /
\sigma_{8,{\rm FIR}} \simeq 1.1$.

In the following we call $\Phi(L)$ the luminosity function of the
catalogue. 
The presence of a flux limit $f_{\rm lim}$ defines a relation between
the distance $r$ and the absolute magnitude $M$ of a galaxy as the
distance at which that galaxy is observed at the flux limit:

\begin{equation}
L_{\rm min}(r) = 4 \pi r^2 f_{\rm lim}
\label{eq:Mlimit}
\end{equation}

\noindent
We define $n(r)$ as the number density of objects in the
flux-limited catalogue at the distance $r$:

\begin{equation}
n(> L_{\rm min}(r) )=\int_{L_{\rm min}(r)}^{+\infty} \Phi(L)dL
\label{eq:selfun}
\end{equation}

\begin{figure*}
\centerline{
\psfig{figure=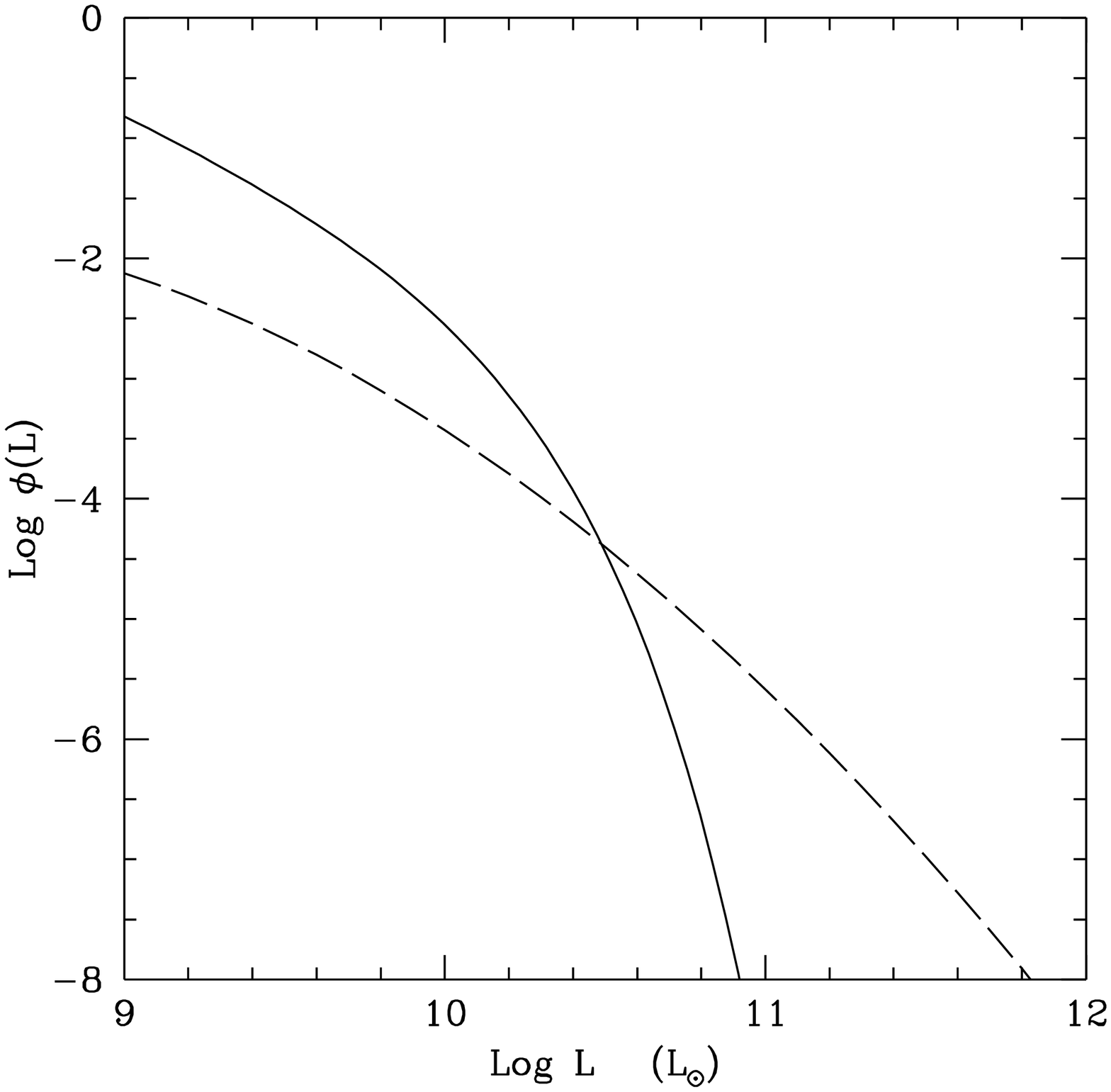,width=9cm}
\psfig{figure=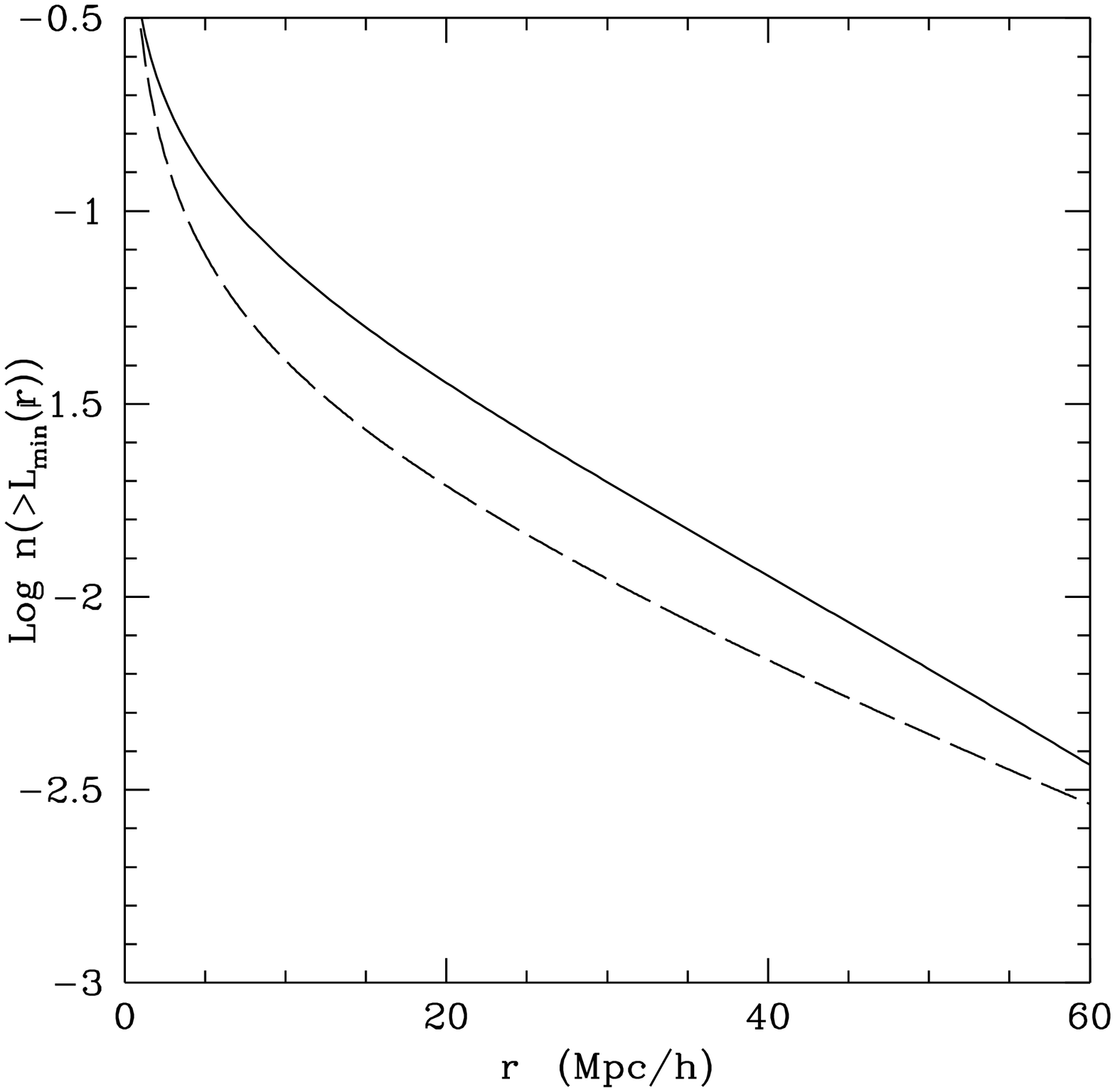,width=9cm}
}
\caption{Left panel shows the Luminosity Function of NOG (continuous
line) and PSCz catalogue (dashed line). Right panel shows the number
density of galaxies of NOG (continuous line) and PSCz catalogue (dashed
line).}
\label{fig:funzioni}
\end{figure*}

\noindent
Given a reference luminosity $L_s$, the selection function $S(r)$ is
defined as the fraction of galaxies more luminous than $L_s$ that are
visible at the distance $r$:

\begin{equation}
S(r)=\frac{\int_{max(L_s,L_{min}(r))}^\infty
\Phi(L)dL}{\int_{L_s}^\infty \Phi(L)dL}
\label{eq:selfun2}
\end{equation}

\noindent
while its inverse $F(r)\equiv 1/S(r)$ gives the number of galaxies
more luminous than $L_s$ present on average at distance $r$ for each
observed galaxy. For the NOG catalogue we use the luminosity function
given by Marinoni (2000); in terms of a Schechter function:

\begin{equation}
\Phi(L)=\Phi^{\ast} \left ( {\frac{L}{L^{\ast}}} \right )^{\alpha}
e^{-L/L^{\ast}}
\label{eq:schecter}
\end{equation}

\noindent
with parameters $\Phi^{\ast} = 0.006 \ Mpc^{-3}(h/0.75)^3$, $M^{\ast} =
-20.6+5\log(h/0.75)$, $\alpha = -1.1$. The luminosity function in the
infrared in given by Saunders et al. (1990). In this work we use the
number density for the PSCz catalogue in the form (Saunders et al.,
2000):

\begin{equation}
n(>L_{min}(r))= \Phi^{\ast}
\frac{(r/r^{\ast})^{1-\alpha}}{[1+(r/r^{\ast})^\gamma]^{\beta/\gamma}}
\label{eq:saunders}
\end{equation}

\noindent
with parameters $\Phi^{\ast} = 0.132 \ h^3 Mpc^{-3}$, $r^{\ast} = 64.6
\ Mpc/h$, $\alpha = -1.1$, $\beta = 3.90$, $\gamma = 1.64$.
Figure~\ref{fig:funzioni} shows all these functions for the NOG and
PSCz catalogues. The higher sampling rate of NOG is evident in this
figure.

\begin{figure}
\centerline{
\psfig{figure=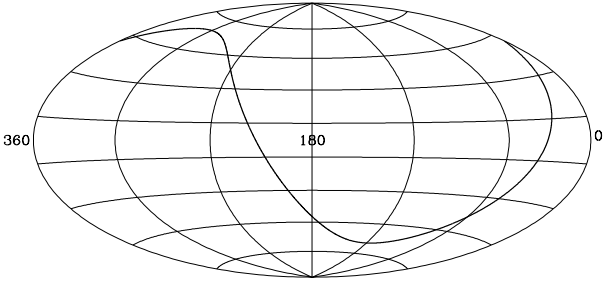,width=9cm}
}
\caption{The figure shows the binning in sky coordinates (galactic
coordinates) employed in the analysis of relative bias between the
catalogues. The thick line represent the celestial equator. Binning in
redshift coordinates is 1200 $km\ s^{-1}$.}
\label{fig:binning}
\end{figure}

The galaxy density fields of the two catalogues have been computed in
different ways, so as to test the robustness of the result.  The
number of optical or IR galaxies has been computed in boxes which are
either sectors of spherical coronae or cubes.  The first choice
corresponds to a binning of galactic coordinates plus redshift; the
binning of sky coordinates is shown in figure~\ref{fig:binning}, while
the binning in redshift is 1200 $km\ s^{-1}$.  The advantage of this
choice is that all boxes are always either inside or outside the
volume covered by both catalogues (i.e. that of NOG), and that it
makes very easy to single out possible problems connected with
inhomogeneities in the selection, like differences between northern
and southern hemispheres.  The second choice corresponds to a binning
of Cartesian coordinates, and allows to keep both the geometry and the
volume of the boxes fixed.  The catalogues have been immersed into a
large box of side 120 \hmpc, with the Galaxy obviously at the centre.
Then the large box has been subdivided into an integer number of small
boxes.  We consider only boxes that overlap for at least 75 per cent
with the NOG volume.  In the following we show the case of boxes of
10 \hmpc.

The density contrast is defined as:

\begin{equation}
\delta=\frac{N_{\rm gal}-N_{\rm ave}}{N_{\rm ave}}
\label{eq:delta}
\end{equation}

\noindent
It has been computed with two different methods.  In the first case,
hereafter method I, we compare the number of galaxies $N_{\rm gal}^I$
found in the box at distance $r$ with the expected one $N_{\rm
ave}^I$, computed as:

\begin{equation}
N_{\rm ave}^I=V \int_{L_{min}(r)}^\infty \Phi(L)dL
\label{eq:metodoI}
\end{equation}

\noindent
where $\Phi(L)$ is the luminosity function of the catalogue and
$L_{\rm min}(r)$ is defined in equation~\ref{eq:Mlimit}. In this case
the shot-noise error associated with the density is:

\begin{equation}
\epsilon(1+\delta)=\frac{\sqrt{N_{\rm gal}^I}}{N_{\rm ave}^I}
\label{eq:erroreI}
\end{equation}

Alternatively (method II) we fix a limit luminosity $L_s$ for
the galaxies, and estimate the number of galaxies with $L>L_s$
present in the box as:

\begin{equation}
N_{\rm gal}^{II} = \sum_{i=1}^n F(r_i)
\label{eq:metodoII}
\end{equation}

\noindent
The density contrast is obtained as in equation~\ref{eq:delta},
where the average number of galaxies this time is:

\begin{equation}
N_{\rm ave}^{II} = V \int_{L_s}^\infty \Phi(L)dL
\label{eq:averageII}
\end{equation}

\noindent
In this case the shot-noise error for the density is (Strauss \&
Willick 1995):

\begin{equation}
\epsilon(1+\delta)=\frac{1}{N_{\rm ave}^{II}} \bigg [ \sum_i F^2(r_i)
\bigg ]^{1/2}
\label{eq:erroreII}
\end{equation}

\begin{figure}
\centerline{
\psfig{figure=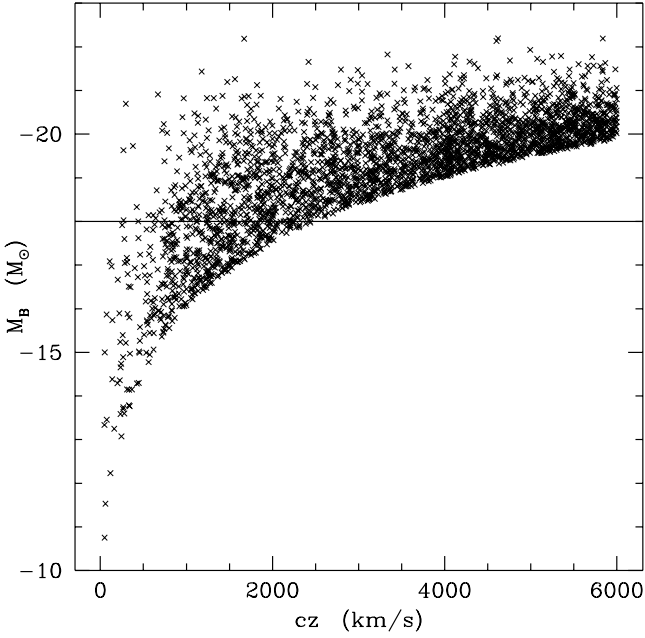,width=9cm}
}
\caption{Absolute Magnitude - Redshift plot for the NOG
catalogue. Objects with absolute magnitude greater than $-18$ are
excluded from the analysis. They represent 15 \% of the total catalogue.}
\label{fig:magni}
\end{figure}

The advantage of method I with respect to method II relies in the
lack of necessity to define a limit luminosity (it is implicitly
defined by the magnitude limit).  In any case, we fix limit
luminosities $L_s$ (and correspondingly a distance $r_s$) for both
catalogues for the following reason.  In the NOG case, we know that
dwarf galaxies trace a different density field with respect to the
luminous ones, and that they are observed only in the inner part of
the NOG volume, where they dominate in number.  In this case, the
density field in the inner part, which is oversampled, would be
dominated by the dwarf population.  This can induce some bias in the
reconstruction.  To avoid this we fix a limit absolute magnitude of
-18.00 and cut both the catalogue and the luminosity function below
this limit.  Figure~\ref{fig:magni} shows the absolute magnitude --
redshift plot for the NOG catalogue; the galaxies excluded are only
16.5 per cent, and all lie within $r_s = 25.00$ \hmpc.  For the PSCz
catalogue, the poor relationship between FIR luminosity and galaxy
mass makes the previous argument inapplicable.  However,
Rowan-Robinson et al. (1991) report an incompleteness of the PSCz
catalogue for the nearest galaxies; following their suggestion we set
$r_s=6$ \hmpc.

\begin{figure}
\centerline{
\psfig{figure=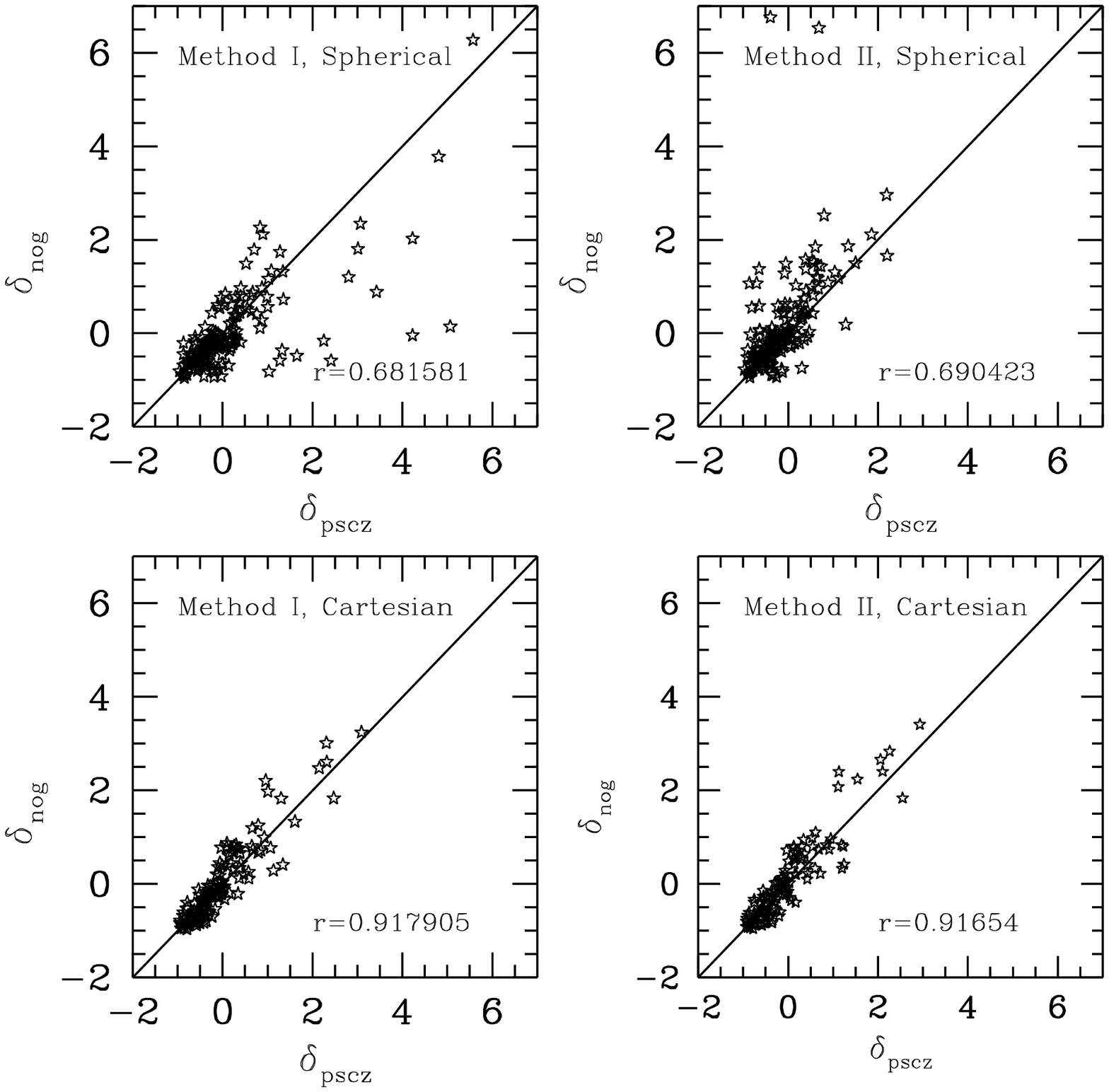,width=9cm}
}
\caption{Correlation between PSCz and NOG density fields in a linear
plane. Methods and volumes employed in the analysis are indicated at
the top of each panel. Rank correlation coefficients are reported at
the bottom of each panel.}
\label{fig:denslin}
\end{figure}

\begin{figure}
\centerline{
\psfig{figure=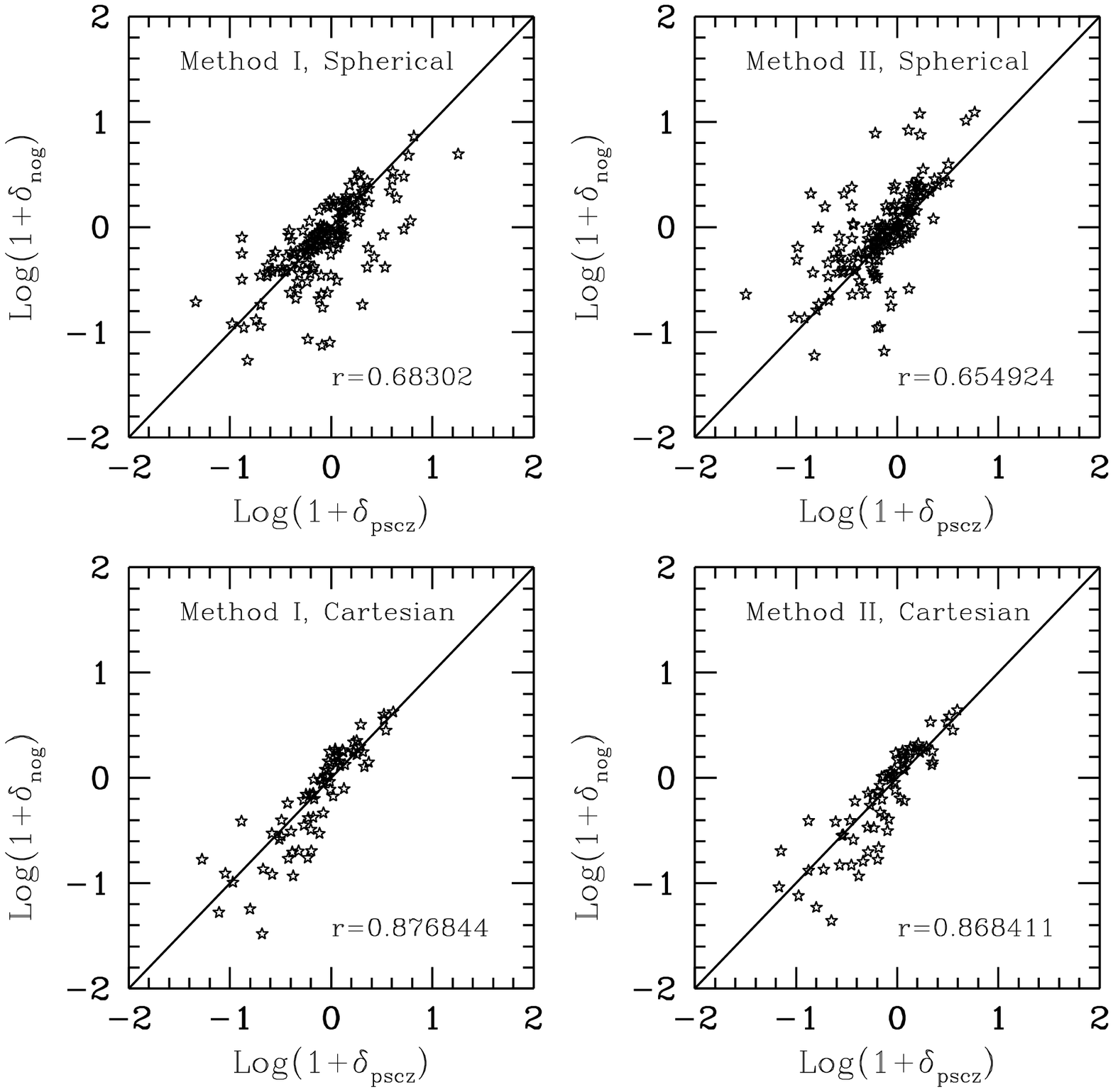,width=9cm}
}
\caption{Correlation between PSCz and NOG density fields in a
bilogaritmic plane. Methods and volumes employed in the analysis are
indicated at the top of each panel. Rank correlation coefficients are
reported at the bottom of each panel.}
\label{fig:denslog}
\end{figure}

Figures~\ref{fig:denslin} and~\ref{fig:denslog} show the resulting
correlation between PSCz and NOG densities, for the two kinds of boxes
and the two methods I and II. Each panel shows the bisector line and
the standard rank correlation coefficient $r$ for the correlation. We
show the correlation both in $\delta$ and in $\log(1+\delta)$ because
it is more convenient to describe a bias relation in the
$\log(1+\delta)$ variable, as the $\delta\ge -1$ constraint is
automatically satisfied.  For all four methods the correlation is very
good, with correlation coefficients exceeding 0.65 in all cases, and
as high as 0.8-0.9 in the case of Cartesian boxes.
The greater scatter found for spherical volumes with respect to
the cartesian ones is due to their different shapes (see section 3.1,
below). This effect could be induced in principle by a scale-dependent
bias. However, we do not find significant variations in the bias
parameter when estimated within cartesian boxes of size from 7.5 to
20$h~{-1}$Mpc. We note that this scale range covers most of the range
probed by the spherical volumes. The quality of fitting is stable with
distance and position on the sky, suggesting that the inhomogeneity of
NOG induces no significant error in the reconstruction of the density
(see also Marinoni 2000).

All the plots have been fit by a linear relation, to determine the bias
parameters.  In particular:

\begin{equation}
\begin{array}{lcl}
\delta_{\rm nog}&=&a_{\rm lin}+b_{\rm lin} \delta_{\rm pscz}
\\ \log(1+\delta_{\rm nog})&=&a_{\rm log}+b_{\rm log}
\log(1+\delta_{\rm pscz}) \\
\end{array}
\label{eq:lineari}
\end{equation}

\noindent
It is worth stressing that the two $a_{\rm lin}$ and $b_{\rm lin}$ or
$a_{\rm log}$ and $b_{\rm log}$ parameters are not independent, as
they are subject to the constraint $\langle \delta \rangle = 0$ for
both density fields.  Moreover, the linear coefficients can be
obtained by the first-order term of the Taylor expansion of the log
relation.  Table~\ref{table:coefficienti} gives the $a$ and $b$
coefficients for all the regressions.  The $b$ bias coefficients are all
consistent, and range from 1.10 to 1.20, with an error of 0.1, with
only one exception, corresponding to a low $r$-value for the
correlation.  The no-bias case is never significantly rejected.  We
conclude that the bias relation between the FIR and the optical galaxy
density fields is:
\begin{equation}
b_{\rm rel} = 1.1 \pm 0.1
\label{eq:bias}
\end{equation}

\begin{table}
\caption{$a$ and $b$ coefficients for all the regressions
(equations~\ref{eq:lineari}). The fits are performed assuming
Poisson errors on both axes. With the simbol $\sigma_r$ we indicate
the scatter of $\delta_{\rm nog}$ around the fit.}
\label{table:coefficienti}
\begin{center}
\begin{tabular}{|c|c|c|c|c|}
\hline Binning & $\delta$ & $b_{lin}$ & $a_{lin}$ & $\sigma_r$ \\
\hline Sky coord. & Met. I & 1.08 $\pm$ 0.10 & 0.04 $\pm$ 0.06 & 1.31 \\
\hline Sky coord. & Met. II & 1.20 $\pm$ 0.10 & 0.17 $\pm$ 0.07 & 1.27
\\
\hline Cartesian coord. & Met. I & 1.17 $\pm$ 0.10 & 0.10 $\pm$ 0.06 &
0.35 \\
\hline Cartesian coord. & Met. II & 1.19 $\pm$ 0.10 & 0.13 $\pm$ 0.07
& 0.37 \\
\\ \hline Binning & $\delta$ & $b_{log}$ & $a_{log}$ & $\sigma_r$ \\
\hline Sky coord. & Met. I & 0.94 $\pm$ 0.09 & -0.01 $\pm$ 0.05 & 0.36
\\
\hline Sky coord. & Met. II & 1.11 $\pm$ 0.11 & 0.10 $\pm$ 0.06 & 0.38
\\
\hline Cartesian coord. & Met. I & 1.16 $\pm$ 0.10 & 0.02 $\pm$ 0.06 &
0.23 \\
\hline Cartesian coord. & Met. II & 1.12 $\pm$ 0.10 & 0.06 $\pm$ 0.06
& 0.23 \\
\hline
\end{tabular}
\end{center}
\end{table}

\noindent
This is consistent with our expectation of $b_{\rm rel}\simeq 1.1$.

We notice that the points that lie most significantly out of the
correlation are those contained in the first bin in redshift ($cz <
1200 \ km\ s^{-1}$). If we exclude these points from the analysis we
find no relevant variation in the values of fit parameters, while the
rank correlation coefficient increases significantly. It is possible
that this behaviour is related to the incompleteness of PSCz in the
nearest region or by the presence of the E-rich Virgo cluster
region. We have also tried to change the number and shape of the
volumes considered, obtaining a set of values for the $a$ and $b$
coefficients consistent with those given in
table~\ref{table:coefficienti}.

An estimate of the $\chi^2$ statistics for these regressions gives in
all cases a high value, with a very low associated probability. We
interpret this as a signature of relative stochastic bias: the two
density fields have an intrinsic scatter beyond that induced by sparse
sampling.  We arbitrarily try to double the Poisson error given by
equations~\ref{eq:erroreI} and \ref{eq:erroreII}, and obtain
acceptable $\chi^2$ values.  This means that the intrinsic scatter of
the two density fields is similar to the shot noise of the
reconstruction. 
In order to quantify the degree of stochastic bias between the
distribution of optical and infrared galaxies, we use the coefficient
$\sigma_b\equiv\sigma_r/\sigma_\delta$, where $\sigma_r$ is the
scatter of $\delta_{\rm nog}$ around the best fit line (see Table 1)
and $\sigma_\delta$ is the variance of the $\delta_{\rm pscz}$.  We
obtain:

\begin{equation}
\sigma_b \sim 0.25\, .
\label{eq:stoca}
\end{equation}

\noindent
The same parameter was used by Dekel \& Lahav (1999) to quantify
the stochastic bias of galaxies with respect to dark matter.
Interestingly, our value is similar to the prediction of Somerville et
al. (2001) for the stochasticity of optical galaxies with respect to
dark matter, which results in their case in the range $0.18-0.35$
(depending on the absolute magnitude cut), with a stronger signal from
optically bright galaxies. Of course this result strongly relies on
the hypothesis that Poisson errors are correct. Furthemore,
non-linearity in the bias relation (that is not observed in our data)
would lead to an overestimate of $\sigma_b$.

\subsection{Filling the optical ZOA and the external regions}

Due to the non-local character of gravity, it is necessary to fill in
the ZOA with a mock galaxy distribution.  Indeed, if the ZOA were
naively treated as a void, it would push all the surrounding matter
aside, thus biasing heavily the reconstruction.  Filling it with
random galaxies, thus neglecting completely the structure present,
would also distort the reconstruction.  To fill in the FIR ZOA we have
used the scheme of Branchini et al. (1999), also used in Monaco et
al. (2000), which consists in interpolating the density field of two
stripes above and below the ZOA.

In the case of NOG, the optical ZOA is so wide that such an
interpolation would result in a poor representation of the true
density field.  Moreover, NOG is limited to $cz<6000\ km\ s^{-1}$,
while PSCz sparsely samples the density field to a distance of
$\sim200-300$ \hmpc, a region which gives an important contribution to
local bulk motion and tides.

It is convenient to use the information provided by the PSCz
catalogue, together with the observed relation of optical and FIR
galaxy density fields, to generate a mock catalogue of optical
galaxies that fills the ZOA and extends to distances larger than 60
\hmpc.  This is done as follows.  The PSCz catalogue has been immersed
into a box of side 250 \hmpc.  Its density field has been computed on
a grid of 128$^3$ points by using the adaptive smoothing scheme
already used by Canavezes et al. (1998), Monaco \& Efstathiou (1999)
and Monaco et al. (2000). In this scheme a Gaussian smoothing kernel
is associated to each galaxy, with a reference radius $R_{\rm ref}$
that satisfies the relation: 

\begin{equation}
\frac{4 \pi}{3}R^{3}_{\rm ref}(r) n(>L_{\rm min}(r))=10
\label{eq:adap}
\end{equation}

\noindent
(so as to have on average 10 galaxies within a distance
$R_{\rm ref}$).  Next, an adaptively refined radius $R_{\rm adap}$ is
computed, such as to contain exactly 10 galaxies within itself.  Then,
the density field is computed by summing over all kernels.  In this
way the density field is computed with constant signal-to-noise.

After this, the volume of the box is divided into small cells, which
are populated by galaxies according to the probability:

\begin{equation}
\begin{array}{lcl}
P&=&V \cdot S_{\rm NOG} \cdot (1+\delta_{\rm NOG}) \\
 &=&V \cdot S_{\rm NOG} \cdot
[(1+\delta_{\rm PSCz})^{b_{\rm rel}}10^{a_{\rm rel}}] \\
\end{array}
\label{eq:complet}
\end{equation}

\noindent
The boxes are kept small enough so that the probability of hosting a
galaxy is always smaller than one: a volume of about 1 square degree
in sky coordinates and 50 $km\ s^{-1}$ in redshift space is small
enough. Once the box is randomly selected to contain a galaxy, the
object is put into a random position within the box.  With this
procedure the mock catalogue follows both the NOG selection function
and the PSCz density field, rescaled to the NOG one through the
relations defined above.

In practice, we do not use the $a$ and $b$ parameters found by the
linear regressions.  The $b_{\rm rel}$ parameter is set either to 1 or
to 1.2, to be able to test the robustness of the results to
uncertainties on the relative bias.  The corresponding $a_{\rm rel}$
parameters are chosen by requiring a good fit of the selection
function for ten different realizations of the filling mock catalogue.
In this way we obtain $a_{\rm rel}=-0.03$ for $b_{\rm rel}=1.2$, while
$a_{\rm rel}=1$ for $b_{\rm rel}=0$. We do not account for
stochasticity in the filling procedure.

\begin{figure*}
\centerline{
\psfig{figure=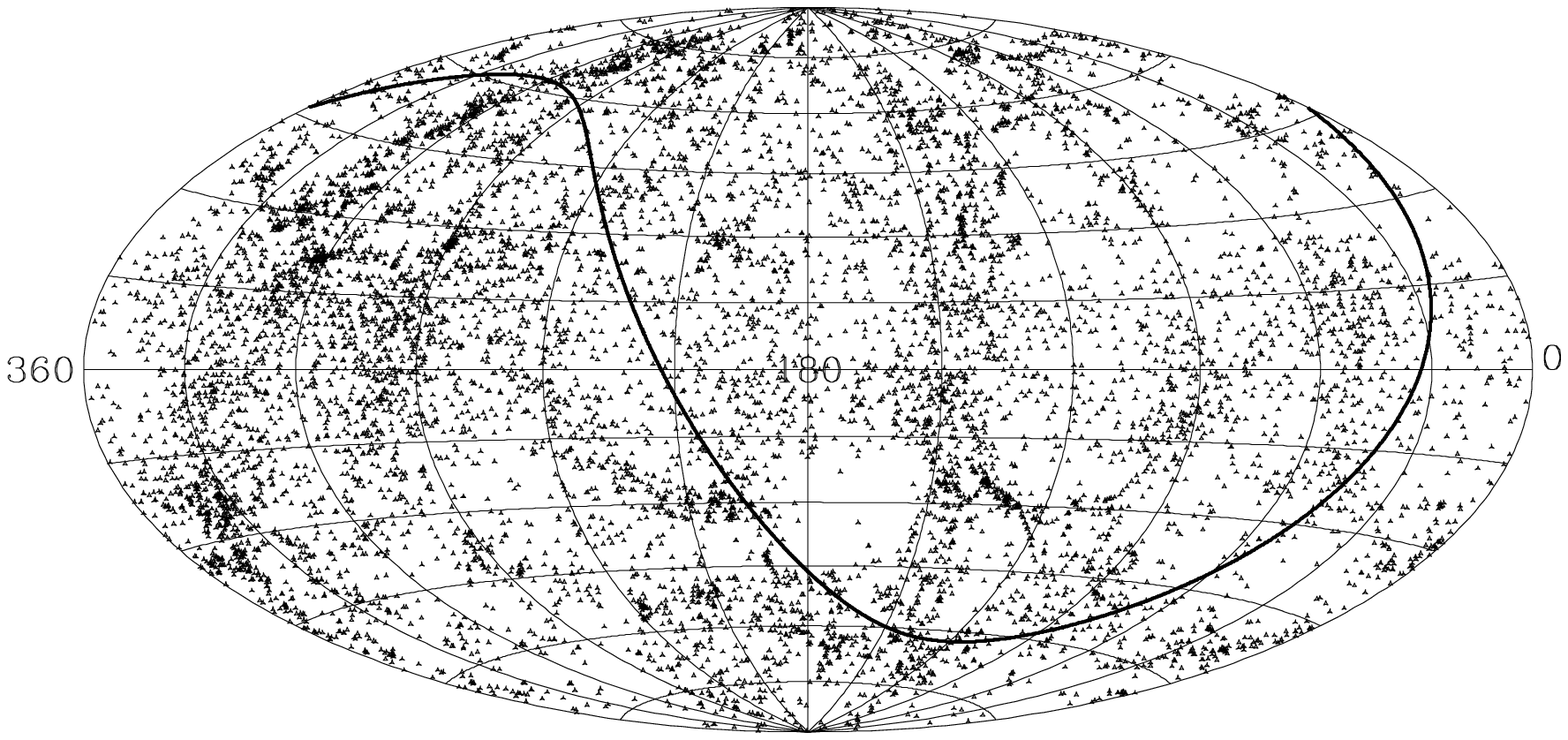,width=9cm}
\psfig{figure=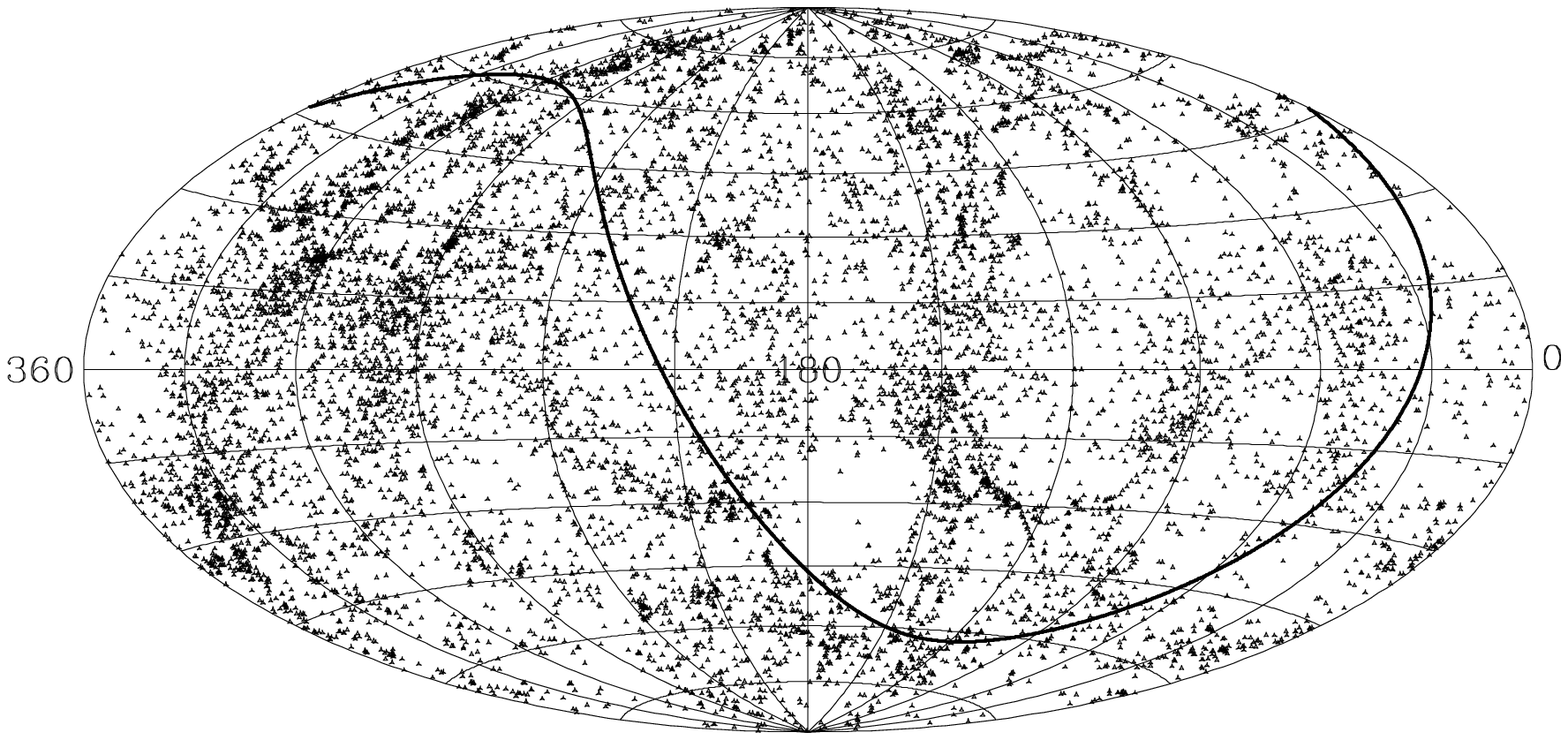,width=9cm}
}
\caption{Completions of the NOG catalogue. Left panel shows the NOG
catalogue with the ZOA filled with the information from PSCz catalogue
in the hypothesis that the relative bias parameter $b_{\rm rel}$ is
1. Right panel shows the NOG catalogue with the ZOA filled with the
information from PSCz catalogue in the hypothesis that the relative
bias parameter $b_{\rm rel}$ is 1.2.}
\label{fig:nogcompl}
\end{figure*}

Figures~\ref{fig:nogcompl} shows the original NOG catalogue completed
with two mock catalogues extracted from the PSCz one, assuming $b_{\rm
rel}=1$ or 1.2.  The structures visible are continued into the ZOA,
although the voids are not as neat as in the real catalogue.
Moreover, the structures in the $b_{\rm rel}=1.2$ case show a slightly
higher contrast.

\section{Tests on simulated catalogues}

The ability of ZTRACE to reconstruct the initial conditions of our
local Universe was demonstrated by Monaco \& Efstathiou (2000).  In
the following we perform further tests aimed to quantify the accuracy
gained by using the denser NOG catalogue with the ZOA filled by the
PSCz and to calibrate the corrections for peak flattening and
smoothing.  In particular, we test the goodness of the reconstruction
of the initial density field in the Fourier space, considering not
only the moduli of the modes (whose average is the power spectrum) but
also their phases, which are mostly responsible for shaping the
observed large-scale structure.

\subsection{Simulated catalogues}

For these tests we perform the ZTRACE reconstruction on galaxy
catalogues extracted from a cosmological N-body simulation.  The
simulation has been run using the Tree gravitational part of the
GADGET code\footnote{http://www.MPA-Garching.MPG.DE/gadget/}
(Springel, Yoshida \& White 2001). We simulate a $\Lambda$CDM model
with $\Omega_m=1-\Omega_\Lambda=0.3$, $\Omega_{bar}=0.019\,h^{-2}$,
$h=0.7$ and $\sigma_8=0.8$, using 256$^3$ dark matter particles within
a box having comoving size of $250\,h^{-1}$Mpc. The Plummer-equivalent
softening length for the computation of the gravitational force has
been set to $\epsilon_{Pl}=30\,h^{-1}$kpc fixed in comoving
units. Initial conditions have been generated at $z_i\simeq 16$ using
the GRAPHIC2 package by Bertschinger (2001).

Mock catalogues are extracted following the selection function either
of the NOG or the PSCz catalogue, and a bias scheme analogous to that
described in section 2 (see equation~\ref{eq:complet}), where the
probability that a particle is selected as a mock galaxy is:
 
\begin{equation}
P=p_b\frac{n(>L_{\rm min}(r))}{n_{\rm max}}
\label{biasass}
\end{equation}

\noindent
Here $n_{\rm max}$ is the mean density of particles in the simulation
and:

\begin{equation}
p_b=a_{\rm sim}(1+\delta)^{(b_{\rm sim}-1)}
\label{biasassbis}
\end{equation}

In this case the density field $\delta$ is the final one traced by all
the particles in the simulation, cic-interpolated on a grid of 128$^3$
and smoothed over 3 \hmpc. We have checked that with our
cosmology the variance of particle counts on boxes of side $L$ is
equal to the variance of the density field Gaussian smoothed on scale
$R$ when $L\sim 3.5 R$; so 3 \hmpc\ Gaussian smoothing is roughly
equivalent to the 10 \hmpc\ boxes shown in figure~\ref{fig:denslin}
and \ref{fig:denslog}.  With this bias scheme mock galaxies are
power-law biased with respect to the dark matter. Again, the $b_{\rm
sim}$ parameter is fixed to some value, while the $a_{\rm sim}$
parameter is chosen so as to reproduce the selection function.

The volume of the NOG catalogue (a sphere of 60 \hmpc) is only 6 per
cent of the volume of the simulation, so that it is safe to extract
ten independent catalogues from the simulation.  We extract four sets
of ten catalogues, all centred on the same set of ten positions
(periodic boundary conditions are assumed).  Two of them are extracted
with the NOG selection function, one unbiased ($b_{\rm sim}=1$,
$a_{\rm sim}=0$) and the other with $b_{\rm sim}=1.2$, $a_{\rm
sim}=-0.03$; two more are selected with the PSCz selection function
and either unbiased or with $b_{\rm sim}=0.8$ and $a_{\rm sim}=0.03$.

After defining a galactic plane (the x-y plane defined by the box) the
optical ZOA and the part with $d>60$ \hmpc\ are taken out from the NOG
catalogues, and then filled with the corresponding simulated PSCz
catalogues following the procedure already outlined in section 2.3:
(i) we compute the density of the corresponding simulated PSCz
catalogues by adaptively smoothing with fixed signal-to-noise; (ii) we
extract a mock catalogue which samples the PSCz density with a given
bias scheme; (iii) with it we fill the optical ZOA and the external
parts. The real PSCz is also subject to a procedure to fill the ZOA,
but we neglect to apply this to the mock catalogues.  Indeed, it was
shown in Monaco \& Efstathiou (1999) that the impact of this filling
procedure is modest, and surely smaller than the corresponding
procedure for the NOG. The relative bias schemes assumed for the
filling procedure are an unbiased one, applied to the pairs of
unbiased NOG and PSCz catalogues, and a biased one, applied either to
the biased NOG and unbiased PSCz, or to the unbiased NOG and
(anti)biased PSCz.  In total, we reconstruct seven sets of ten
simulated catalogues; these are listed in
table~\ref{table:simulazioni}.

\begin{table}
\caption{Simulated Catalogues}
\label{table:simulazioni}
\begin{center}
\begin{tabular}{|c|c|}
\hline Catalogue & $b_{\rm sim}$ \\ 
\hline
NOG unbiased & $1$ \\
NOG biased & $1.2$ \\ 
PSCz unbiased & $1$ \\
PSCz biased & $0.8$  \\
\hline
Catalogue & $b_{\rm rel}$ \\
\hline 
NOG unbiased filled with PSCz unbiased & $1$ \\
NOG biased filled with PSCz unbiased & $1.2$ \\
NOG unbiased filled with PSCz biased & $1.2$ \\
\hline
\end{tabular}
\end{center}
\end{table}

\begin{figure}
\centerline{
\psfig{figure=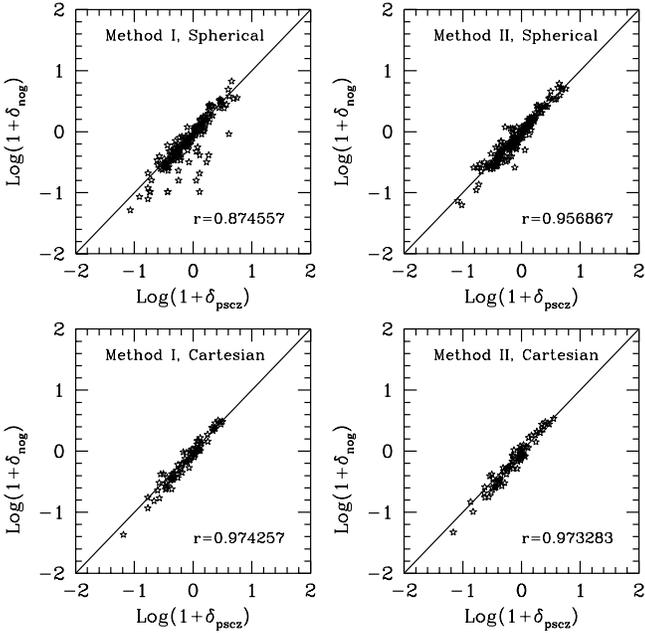,width=9cm}
}
\caption{As in figure~\ref{fig:denslog} for one simulated pair of
catalogues: NOG with $b_{\rm sim}=1.2$, PSCz unbiased.}
\label{fig:simdenslog}
\end{figure}

As a first check we repeat the procedure described in section
2.2, i.e. we divide the volume into subvolumes and compute the
``galaxy'' density contrasts for pairs of simulated catalogues.
Figure~\ref{fig:simdenslog} shows the bilogarithmic correlation of
density contrasts for one realization of unbiased PSCz and biased
($b_{\rm sim}=1.2$) NOG.  As in figure~\ref{fig:denslog}, spherical
volumes show a greater scatter than cartesian ones, especially when
Method I is used.  As anticipated above, this confirms that the
variation in scatter is mostly connected to the volume geometry.  The
outcoming slopes of the relations are for the four cases $b_{\rm
rel}=1.22 \pm 0.06$ (Method I, Spherical), $b_{\rm rel}=1.18 \pm 0.05$
(Method II, Spherical), $b_{\rm rel}=1.11 \pm 0.06$ (Method I,
Cartesian) and $b_{\rm rel}= 1.06 \pm 0.03$ (Method II, Cartesian).
This leads to an average estimate of $b_{\rm rel}=1.14 \pm 0.06$,
consistent with the true $b_{\rm rel}=1.2$ value, although we notice
(also in the other realizations) a slight tendency to underestimation
of $b_{\rm rel}$.  Moreover the $\chi^2$ tests give reasonable values
with high associated probability, confirming that the scatter in this
case is due to Poisson errors; this supports our interpretation of the
scatter in the data as stochasticity of bias, although with all the
caveats highlighted above.

\subsection{ZTRACE reconstruction}

Here we outline briefly the main steps of the ZTRACE reconstruction.
This applies to the simulated catalogues as well as to the real ones.
Full details are given in Monaco \& Efstathiou (1999).

First the relaxed groups are collapsed, so as to remove the ``finger
of God'' features present in the redshift space; this is known to
improve the reconstruction (Gramann et al. 1994).  In all cases but
the real NOG catalogue, to select the groups we use a standard
friends-of-friends code with linking lengths 0.5 and 3 \hmpc\
respectively tangential and along the line of sight. In the NOG case
we use the groups found by Giuricin et al. (2000). This operation is
obviously carried over the catalogues, and not over the filling of the
ZOA.

Second, the catalogues are smoothed with the same adaptive scheme as
that described in section 2.3.  This time the reference smoothing
radius is kept constant to $R_{\rm max}$ within a distance $d_{\rm
max}$, such as to have $4\pi/3\ R_{\rm max}^3 n(>L_{\rm min}(d_{\rm
max}))=10$, and then changed as in equation~\ref{eq:adap} so as to
keep the signal-to-noise constant.
The adaptive refinement is done so as to have $4\pi/3\ R_{\rm ref}^3
n(>L_{\rm min}(r))$ galaxies within $R_{\rm adap}$.  In this way the
inner part ($r<d_{\rm max}$) is adaptively smoothed with constant
reference smoothing radius, while the outer parts are smoothed with a
reference smoothing radius which increases with distance. For choices
of $d_{\rm max}=30$ or 60 \hmpc\ we have $R_{\rm max}=$ 3.17 and 5.57
\hmpc\ for NOG, 3.83 and 5.99 \hmpc\ for PSCz.

Third, the density field is given as an input to the ZTRACE code, thus
obtaining a reconstruction of the initial conditions.  
These are given in terms of the quantity:

\begin{equation}
\delta_l=\frac{\delta({\bf x},t_i)}{D(t_i)}
\label{eq:deltal}
\end{equation}

\noindent
where $\delta({\bf x}, t_i)$ is the density contrast field at a very
early time $t_i$, when linear theory holds, and $D(t_i)$ is the linear
growing factor at $t_i$.  This quantity, called linear density
contrast, is constant in linear theory, and is equal to the linear
extrapolation of the initial conditions to $z=0$ (where $D(t)=1$ by
normalization).  At variance with Monaco \& Efstathiou (1999), the
initial conditions are reconstructed on a grid of a 128$^3$ (instead
of 64$^3$)\footnote{This way the smoothing at the centre of
coordinates, necessary to ensure convergence of the iteration, affects
only the density field within $\sim 5$ \hmpc.}.

\subsection{Fixing the flattening of the peaks}

The ZTRACE algorithm is known to preserve the statistics of the
initial conditions for density contrasts lower than $\delta_l \sim
1$, but it flattens the high-density peaks.  This is due to the fact
that high peaks are sites of multi-streaming, a regime in which any
reconstruction based on the mildly non-linear evolution fails; ZTRACE
assumes that the density field is always in single-stream regimes.
This distortion induces some degree of non-Gaussianity in the
reconstruction that must be removed.

To achieve this we compute the PDF of the reconstructed initial
conditions for all the ten reconstructions of a given set of simulated
catalogues, and from these the average initial PDF.  Galaxy bias is
known to induce distortions in the initial PDF (Monaco et al. 2000),
but the effect is negligible for the moderate bias schemes used here.
To decrease the scatter in the PDF from one realization to the other,
we rescale all the PDFs to zero mean and unit variance, use the median
instead of the mean and compute the variance using only the
underdensities, were the density contrast is best reproduced.  The
reconstructed initial PDF is found to be always consistent with
Gaussian for $\delta\le 0$.  For $\delta>0$ we choose a parametric
transformation $\delta'=f(\delta)$ such that the PDF of the $\delta'$
variable is Gaussian to a good approximation.  For this transformation
we use a branch of hyperbola, subject to the constraint of vanishing
at $\delta=0$ with first derivative equal to unity (so as to join
smoothly the bisector line), and to have a flat asymptote at
$\delta\rightarrow \infty$:

\begin{equation}
y= \left\{ \begin{array}{ll} ax-\sqrt{c_1^2x^2+c_2^2(c_1-1)x+c_2^2}+c_2
& {\rm if} \  x>0 \\
x & {\rm if} \  x<0 \\
\end{array} \right.
\label{eq:raddrizza}
\end{equation}

\noindent
The $c_1$ and $c_2$ parameters determine respectively the curvature of
the function and the level of the asymptote.  The parameters have been
found for each reconstructed PDF; they are listed in
table~\ref{table:picchi}.

\begin{table}
\caption{Parameters of the correction for the flattening of the peaks}
\label{table:picchi}
\begin{center}
\begin{tabular}{|c|c|c|c|c|}
\hline Catalogue & \multicolumn{2}{|c|}{$d_{\rm max}=60 Mpc/h$} & 
\multicolumn{2}{|c|}{$d_{\rm max}=30 Mpc/h$} \\
\hline 
 & $c_1$ & $c_2$ & $c_1$ & $c_2$ \\ 
\hline
NOG unbiased & 1.03 & 6.50 & 0.95 & 4.20 \\
NOG biased & 1.00 & 4.70 & 0.90 & 3.70 \\ 
PSCz unbiased & 1.05 & 4.80 & 0.97 & 4.95 \\
PSCz biased & 1.05 & 6.00 & 1.00 & 4.70 \\
NOG unbiased filled & 1.10 & 5.10 & 1.10 & 3.50\\ 
with PSCz unbiased & & & & \\
NOG unbiased filled & 1.04 & 5.30 & 0.95 & 3.90\\
with PSCz biased & & & & \\
NOG biased filled & 1.05 & 4.00 & 0.95 & 4.00 \\
with PSCz unbiased & & & & \\
\hline
\end{tabular}
\end{center}
\end{table}

\begin{figure*}
\centerline{
\psfig{figure=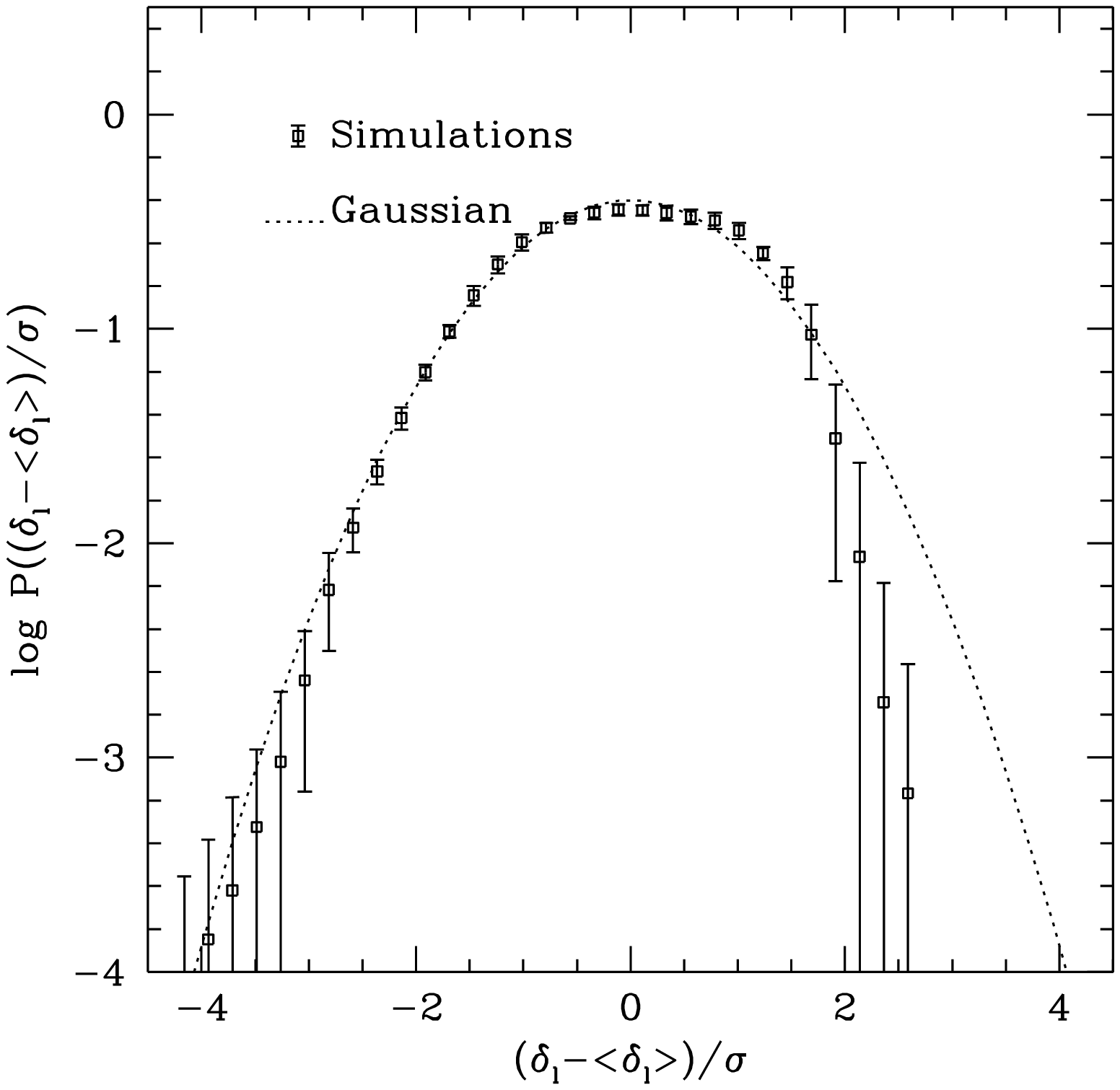,width=9cm}
\psfig{figure=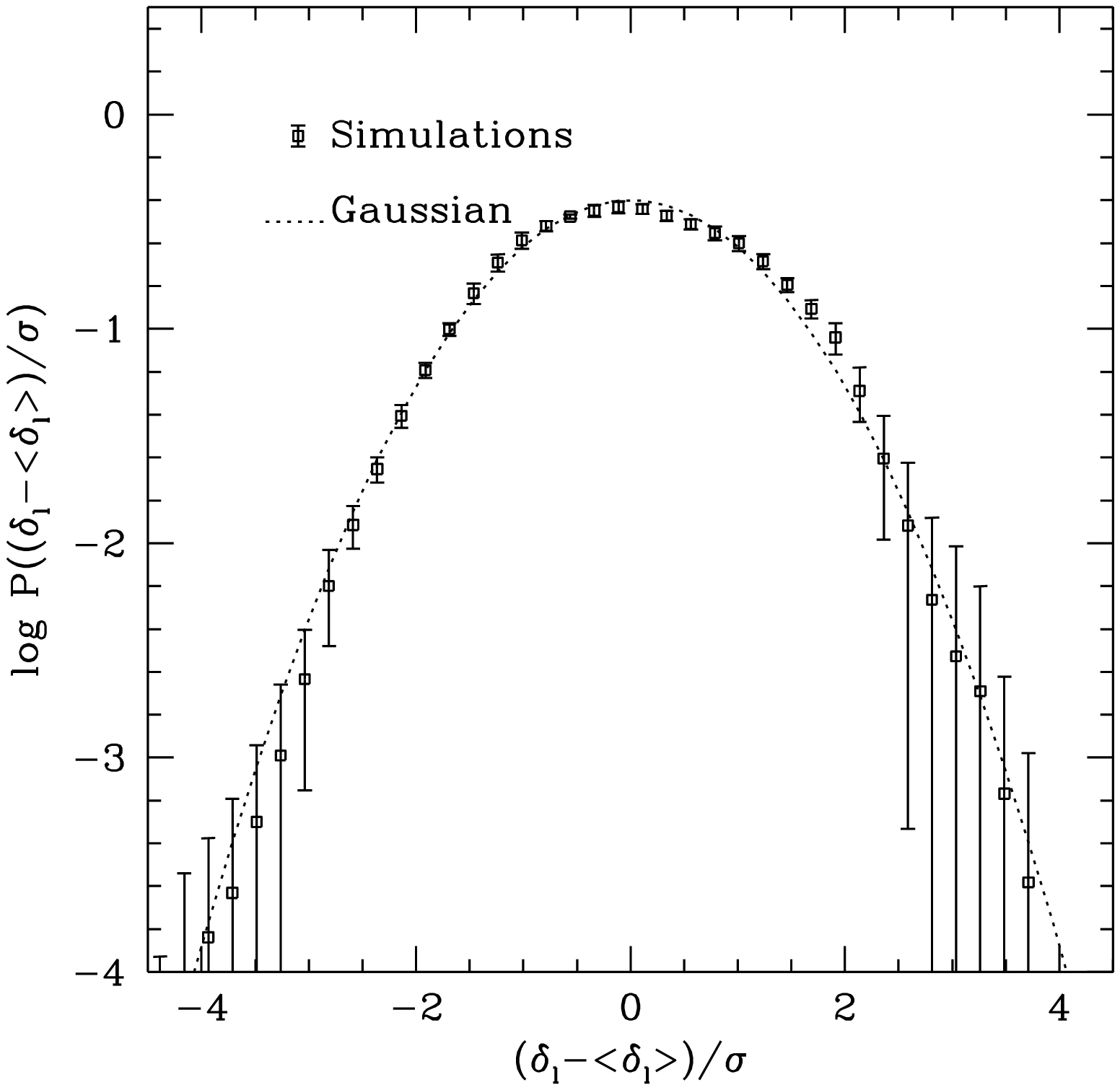,width=9cm}
}
\caption{Average reconstructed PDFs.  Left panel shows the average
reconstructed PDF for ''NOG-unbiased filled with PSCz--unbiased''
simulated catalogues. The non-Gaussianity is due to the flattening of
high-density peaks. Right panel shows the same PDFs after the
correction of peak flattening (with coefficients $c_1 = 1.10$, $c_2 =
5.10$; see table~\ref{table:coefficienti}). In both panels dotted
lines show a Gaussian of zero mean and unit variance.}
\label{fig:picchia}
\end{figure*}

\begin{figure}
\centerline{
\psfig{figure=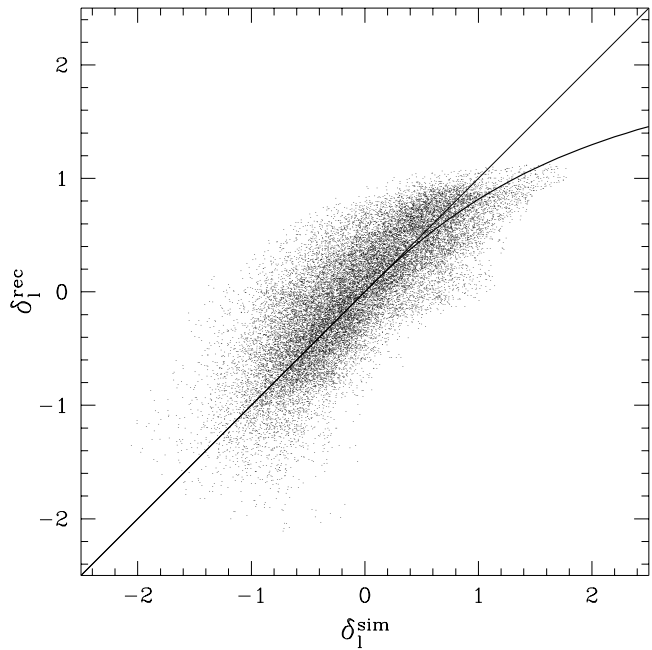,width=9cm}
}
\caption{Scatter-plot of true and reconstructed linear densities for
one of the simulated catalogues (NOG unbiased filled with PSCz
unbiased), with the best-fit transformation.}
\label{fig:picchib}
\end{figure}

Figure~\ref{fig:picchia} shows, as an example, the average
reconstructed PDF for the unbiased NOG catalogues with the ZOA filled
with the unbiased PSCz density fields, before and after the fix for
peak flattening.  Figure~\ref{fig:picchib} shows the scatter-plot of
true and reconstructed density for one of the simulated catalogues,
with the best-fit transformation superposed.

Finally, we notice that our procedure is conceptually different
from the Gaussianization technique of Weinberg (1992), where a
Gaussianity of the PDF is directly forced on a linearly reconstructed
density field.  We apply our ``Gaussianization'' procedure only on the
high-density peaks (the $\delta<0$ density field is completely
untouched) to remove the already analized effect of peak flattening.

\subsection{Accuracy of the reconstruction in the Fourier space}

Due to the inhomogeneous nature of the smoothing performed on the
original catalogues, the initial conditions are not recovered down to
a scale which is constant in space.  Then, to test the reconstruction
in the Fourier space without mixing information from different scales
we pad to zero both the true and the reconstructed linear density
fields beyond $d_{\rm max}$.  We also pad in some cases the ZOA, to
test the reconstruction in the region which is not subject to the
filling procedure.  This padding of course induces some phase
correlation at large scales.  To be sure that any positive correlation
of moduli and phases that we find is due to a true signal, we compare
the reconstructed linear density field also with the true one centred
in a point which is half the box size away from the centre of the
simulated catalogue considered.  In this case the reconstructed and
true initial conditions are completely independent, and any
correlation must be ascribed to the padding procedure.  These initial
conditions will be referred to as "random" in the following.

The true, reconstructed and random initial conditions have been
FFT-transformed, and the modes $\delta_{\bf k}$ have been compared one
by one in terms of modulus $|\delta_{\bf k}|$ and phase $\varphi_{\bf
k}$.  To quantify the agreement the modes have been binned into
intervals of $k$, and for each bin the following statistics have been
computed:

\begin{eqnarray}
S_{\rm module} &=& \sqrt{\left\langle \left[ \frac{|\delta_{\bf k}^{\rm sim}|^2 
-|\delta_{\bf k}^{\rm rec}|^2}{P^{\rm sim}(k)} \right]^2 \right\rangle } \\
S_{\rm phase} &=& \langle (\varphi_{\bf k}^{\rm sim}-\varphi_{\bf k}^{\rm rec})^2
\rangle
\label{eq:statistica}
\end{eqnarray}

\begin{figure*}
\centerline{
\psfig{figure=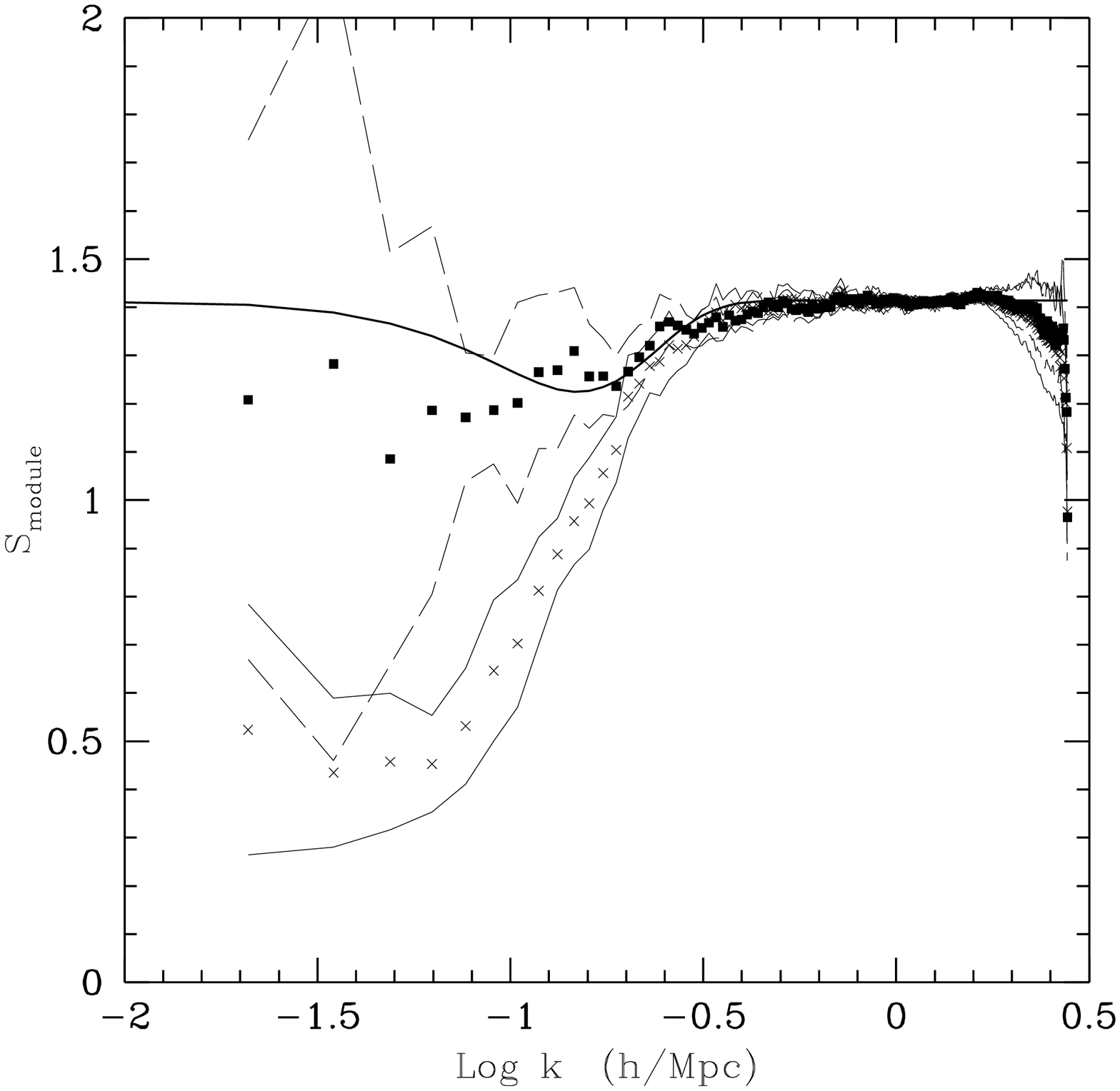,width=9cm}
\psfig{figure=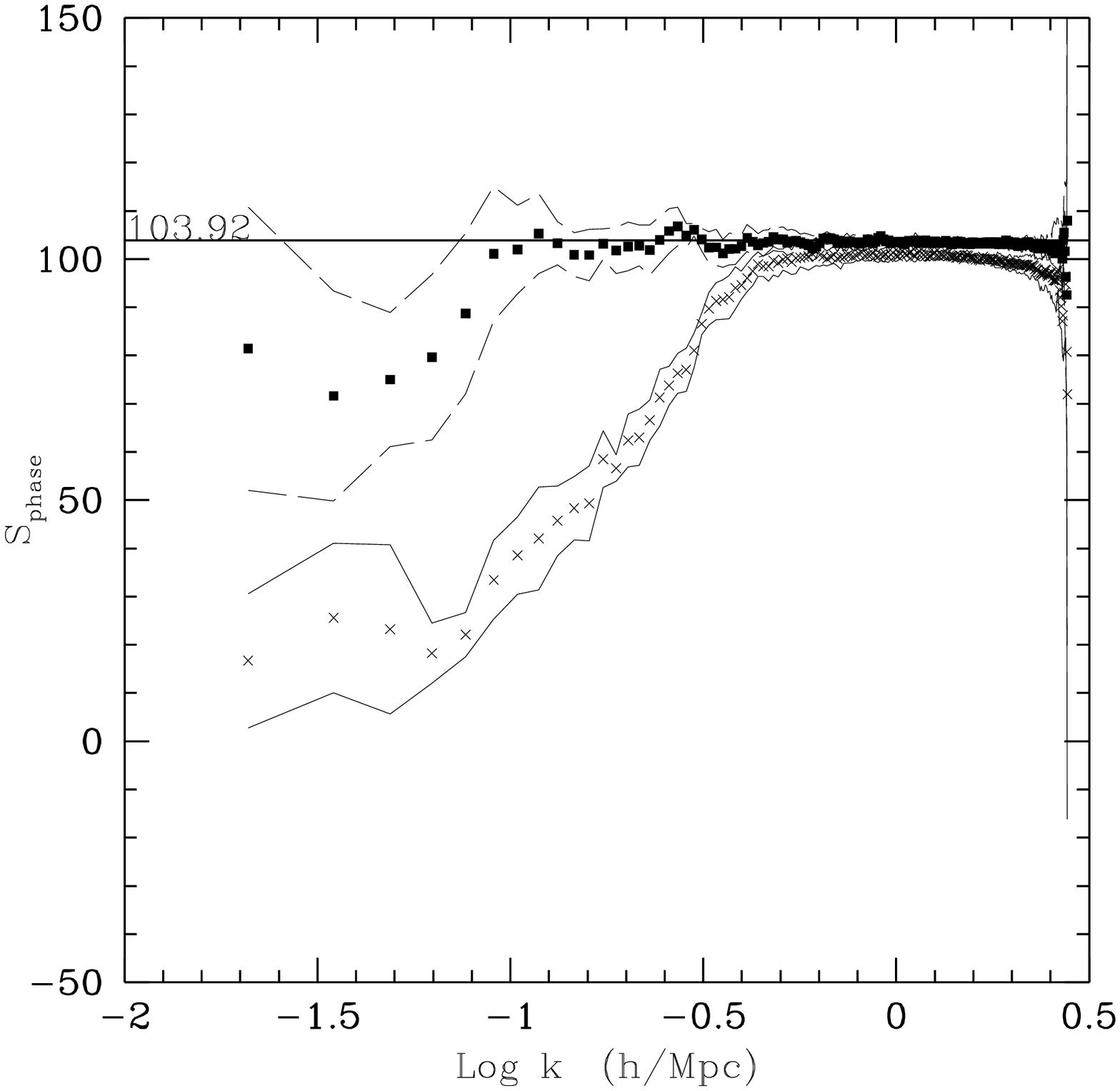,width=9cm}
}
\caption{Statistic of reconstruction. The left panel shows $S_{\rm
module}$. The right panel shows $S_{\rm phase}$. The curves
are averages of the statistics over the ten realizations. Cross points
refer to the comparison of true and reconstructed fields, square
points to the comparison of random and reconstructed ones. Continuous
lines represent the 1-$\sigma$ scatter from the average value for the
comparison of true and reconstructed fields, dashed lines for the
comparison of random and reconstructed ones.  The thick lines show the
expectation value of the statistics.}
\label{fig:stat}
\end{figure*}

Figure~\ref{fig:stat} shows as an example these statistics for the unbiased NOG
catalogues filled with unbiased PSCz's. The curves are averages of the
statistics over the ten realizations, the 1-$\sigma$ scatter is also
shown. Cross points refer to the comparison of true and reconstructed
fields, square points to the comparison of random and reconstructed
ones. The thick lines show the expectation for two completely uncorrelated
fields, which for $S_{\rm module}$ is:

\begin{equation}
\begin{array}{lcl}
S_{\rm module}^{\rm uncor} &=& 2\frac{(P^{\rm sim}(k)^2+P^{\rm
rec}(k)^2 -P^{\rm sim}(k) P^{\rm rec}(k))}{P^{\rm sim}(k)^2}\\
\\ &\simeq& 2\left[\left(1-\exp^{-k^2R^2}\right)^2 + \exp^{-k^2R^2}\right]\\
\end{array}
\label{eq:randmod}
\end{equation}

\noindent
(the last passage is done assuming $P^{\rm rec}\simeq P^{\rm sim}
\exp^{-k^2R^2}$, which is approximately true).  For the phases we
obtain:

\begin{equation}
\begin{array}{lcl}
S_{\rm phase}^{\rm uncor} &=& \int d\phi_1 \int
d\phi_2 \frac{1}{2 \pi^2} (\phi_1-\phi_2)^2\\
\\
 &\simeq& 103.92\\
\end{array}
\label{eq:randphase}
\end{equation}

\noindent
The integral is computed taking into account that $-\pi \le
(\phi_1-\phi_2) \le \pi$.

The first statistics quantifies the correlation of the fluctuations of
the modules of the modes around the average value (i.e. the power
spectrum), the second statistics quantifies the correlation of phases.
From figure~\ref{fig:stat} it is clear that both phases and
fluctuations of modules are significantly reconstructed at large
scales, while the correlation tends to the random value at large
$k$-values.  The transition happens at a $k$-value similar to $1/R$,
so that in this case the worsening of the agreement is mostly due to
smoothing.  Finally, phases are apparently reconstructed more
accurately than the fluctuations of the modules.

It is important to quantify the $k$-value at which the correlation is
lost.  We have tried many possible definitions for this $k_{\rm lim}$
quantity, mostly based on the agreement between the phases, which we
consider more important than the fluctuations of the modules.  Namely,
we have defined $k_{\rm lim}$ as the value at which the correlated
curve is for the first time within (or for the last time beyond)
2$\sigma$ of the uncorrelated curve, with $\sigma$ that relative of
either of the two curves, or the value at which the average correlated
curve is at a fixed fraction (say 90 per cent) of the value relative
to uncorrelated fields (equation~\ref{eq:randphase}).  All these
definitions give similar values for $k_{\rm lim}$.  In particular, the
last definition is the most stable one, as it is not based on the
noisy curves relative to the scatter.  In the following only the
results based on the last definition will be presented.

\begin{table*}
\caption{Results for $k_{lim}$; $d_{max}=60 Mpc/h$}
\label{table:tabellona60}
\begin{center}
\begin{tabular}{|c|c|c|c|c|c|}
\hline 
Catalogues & $R_{\rm max} (Mpc/h)$ & $k_{\rm lim} (h/Mpc)$ & $k_{\rm
lim} R_{\rm max}$ & $k_{\rm lim} (h/Mpc)$ & $k_{\rm lim} R_{\rm max}$ \\
 & & & & ZOA padded & ZOA padded \\
\hline PSCz unbiased & 5.99 & 0.36 & 2.16 & -- & -- \\
\hline PSCz biased & 5.99 & 0.37 & 2.22 & -- & -- \\
\hline NOG unbiased & 5.57 & 0.38 & 2.12 & -- & -- \\
\hline NOG biased & 5.57 & 0.38 & 2.12 & -- & -- \\
\hline NOG unbiased filled & & & & & \\ 
with PSCz unbiased & 5.57 & 0.37 & 2.06 & 0.37 & 2.06 \\
\hline NOG unbiased filled & & & & & \\
with PSCz biased & 5.57 & 0.36 & 2.01 & 0.37 & 2.06 \\
\hline NOG biased filled & & & & & \\
with PSCz unbiased & 5.57 & 0.34 & 1.89 & 0.36 & 2.01 \\
\hline
\end{tabular}
\end{center}
\end{table*}

\begin{table*}
\caption{Results for $k_{lim}$; $d_{max}=30 Mpc/h$}
\label{table:tabellona30}
\begin{center}
\begin{tabular}{|c|c|c|c|c|c|}
\hline 
Catalogues & $R_{\rm max} (Mpc/h$ & $k_{\rm lim} (h/Mpc)$ & $k_{\rm
lim} R_{\rm max}$ & $k_{\rm lim} (h/Mpc)$ & $k_{\rm lim} R_{\rm max}$ \\
 & & & & ZOA padded & ZOA padded \\
\hline PSCz unbiased & 3.83 & 0.38 & 1.45 & -- & -- \\
\hline PSCz biased & 3.83 & 0.40 & 1.53 & -- & -- \\
\hline NOG unbiased & 3.17 & 0.41 & 1.30 & -- & -- \\
\hline NOG biased & 3.17 & 0.41 & 1.30 & -- & -- \\
\hline NOG unbiased filled & & & & & \\ 
with PSCz unbiased & 3.17 & 0.38 & 1.20 & 0.40 & 1.27 \\
\hline NOG unbiased filled & & & & & \\
with PSCz biased & 3.17 & 0.38 & 1.20 & 0.38 & 1.20 \\
\hline NOG biased filled & & & & & \\
with PSCz unbiased & 3.17 & 0.38 & 1.20 & 0.38 & 1.20 \\
\hline
\end{tabular}
\end{center}
\end{table*}

Table~\ref{table:tabellona60} and~\ref{table:tabellona30} give the
resulting $k_{\rm lim}$ for the seven sets of simulated catalogues,
the two choices of $d_{\rm max}$ and $R_{\rm max}$ and the two cases
of no padding or padding of the ZOA.  For reference, we give also the
quantity $k_{\rm lim} R_{\rm max}$, which is a good indicator of the
goodness of the reconstruction.  We conclude that: (i) for $d_{\rm
max}=60$ \hmpc\ the reconstruction is valid up to $k_{\rm lim} R_{\rm
max}\simeq 2$, i.e. at scales as small as half of the (Gaussian)
smoothing radius; (ii) in the ideal case of no ZOA and no bias, the
denser sampling of NOG with respect to PSCz gives a tiny improvement
to the reconstruction in terms of $k_{\rm lim}$, with a decreasing
$k_{\rm lim} R_{\rm max}$; (iii) the correction of the ZOA makes the
advantage of NOG almost vanish; (iv) the gain in $k_{\rm lim}$
obtained by smoothing at a smaller scale (with $d_{\rm max}=30$ \hmpc)
is modest, corresponding to a decrease of $k_{\rm lim} R_{\rm max}$,
which confirms that scales smaller than $\simeq$ 5 \hmpc\ are
dominated by highly non-linear dynamics; (v) in the case of biased
catalogues the reconstruction works to a smaller $k_{\rm lim}$ value,
because the onset of non-linearity is anticipated, while the opposite
is true in the case of antibiased catalogues\footnote{It is fair to
recall here that ZTRACE works always in the hypothesis of no bias, so
that a biased catalogue gives a more non-linear density field.}.

\subsection{The distribution of the phases}

It is useful to check the Gaussianity of the reconstructed initial
conditions directly in terms of distribution of the phases.

\begin{figure}
\centerline{
\psfig{figure=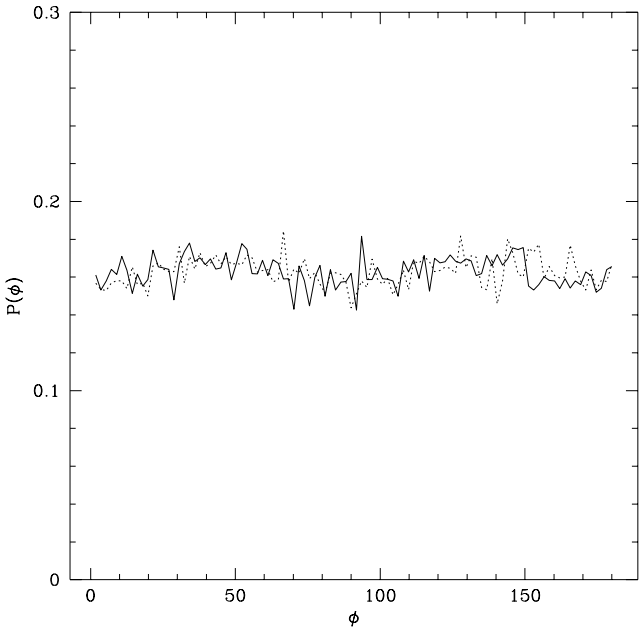,width=9cm}
}
\caption{Distribution of the phases of true (dotted lines) and
reconstructed (continuous lines) initial conditions for $k<k_{\rm
lim}$. The distributions are averages over 10 simulated catalogues. We
show the case for unbiased NOG catalogues filled with unbiased PSCz.}
\label{fig:fasidist}
\end{figure}

Figure~\ref{fig:fasidist} shows, for the unbiased NOG catalogues
filled with unbiased PSCz, the average distribution of phases for
$k<k_{\rm lim}$, together with the average distribution of the phases
of the true initial conditions padded beyond $d_{\rm max}$.  We have
checked that both distributions are consistent with flat by
FFT-transforming them and comparing their spectrum with a white-noise
one.  We conclude that the flatness of the distribution of the
reconstructed phases is preserved by the reconstruction. The same
conclusions hold for the other cases.

\subsection{Correction of the power spectrum for smoothing}

\begin{figure}
\centerline{
\psfig{figure=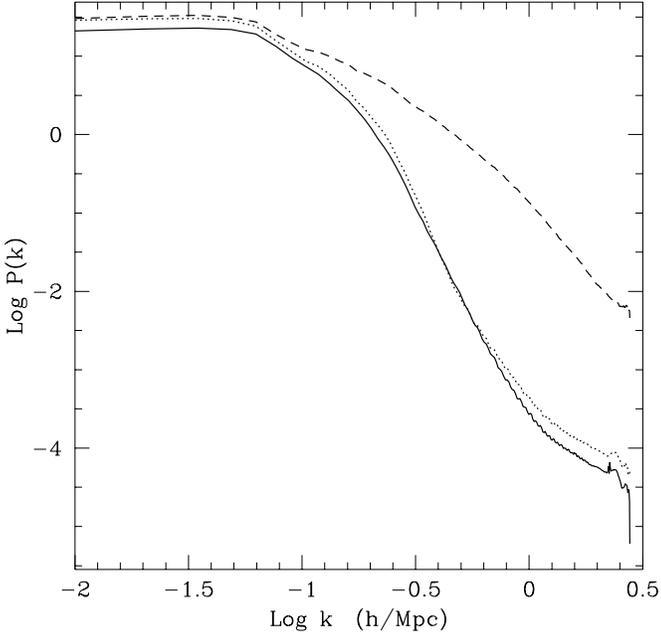,width=9cm}
}
\caption{Power spectra. Dashed line refers to the spectrum of true
padded initial density field; dotted line refers to the spectrum of
true padded initial density field smoothed over 6 $Mpc/h$; continuous
line refers to the average spectrum of reconstructed initial density
fields over 10 simulated catalogues. We show the case for unbiased NOG
catalogues filled with unbiased PSCz.}
\label{fig:spettri}
\end{figure}

Figure~\ref{fig:spettri} shows the power spectra of the padded true,
padded true smoothed (over $R_{\rm max}$) and reconstructed initial
density fields averaged over the ten realizations in the case of
unbiased NOG catalogues filled with unbiased PSCz. The reconstructed
power spectrum is similar to the smoothed one, confirming that the
adaptive smoothing scheme adopted by ZTRACE is very similar to
smoothing in the (Lagrangian) space of initial conditions.  Similar
results are obtained in other cases.  However, the small discrepancies
visible in the figure and present also in the other cases are
sometimes larger than the typical error associated to the power
spectrum (i.e. the rms over the ten realizations).

To restore the power lost by smoothing, it is possible as a first
approximation to simply multiply the FFT-transformed field by the
inverse of the Gaussian smoothing kernel, i.e. by $\exp[(k R_{\rm
max})^2/2]$.  This turns out to be a good approximation for the two
NOG catalogues.  In the PSCz case the correction is improved by
multiplying the Gaussian kernel by the ratio of the reconstructed and
true (smoothed) padded power spectra, as those shown in
figure~\ref{fig:spettri}.  To check the accuracy of these corrections
we discard from the FFT-transformed reconstructions all the modes with
$k>k_{\rm lim}$.  The density fields are then completed with Gaussian
random fields with power spectra truncated below $k_{\rm lim}$.  The
so-completed fields are then transformed them back to the real space,
padded beyond $d_{\rm max}$ and FFT-transformed again to compute the
power spectra.  We obtain smooth spectra in all the three cases,
without any jump at $k_{\rm lim}$.

\section{Results}

The three catalogues (PSCz and the two NOG completions with either
$b_{\rm rel}=1$ or 1.2) have been processed as described in section 3;
for sake of clarity we sum up the procedure again.  (i) The relaxed
groups have been collapsed to spheroids; (ii) the catalogues have been
adaptively smoothed, with reference smoothing radius kept constant
within either 60 or 30 \hmpc; (iii) the density fields have been
processed by ZTRACE to generate the initial conditions; (iv) the
flattening of the peaks has been fixed as described in section 3.3; (v)
the power lost by smoothing has been restored as described in section
3.6.

\subsection{The PDF of the initial conditions from the NOG catalogue}

It is interesting to check the Gaussianity of the initial conditions
reconstructed from the NOG catalogue, as done by Monaco et al. (2000)
for the PSCz.  

\begin{figure*}
\centerline{
\psfig{figure=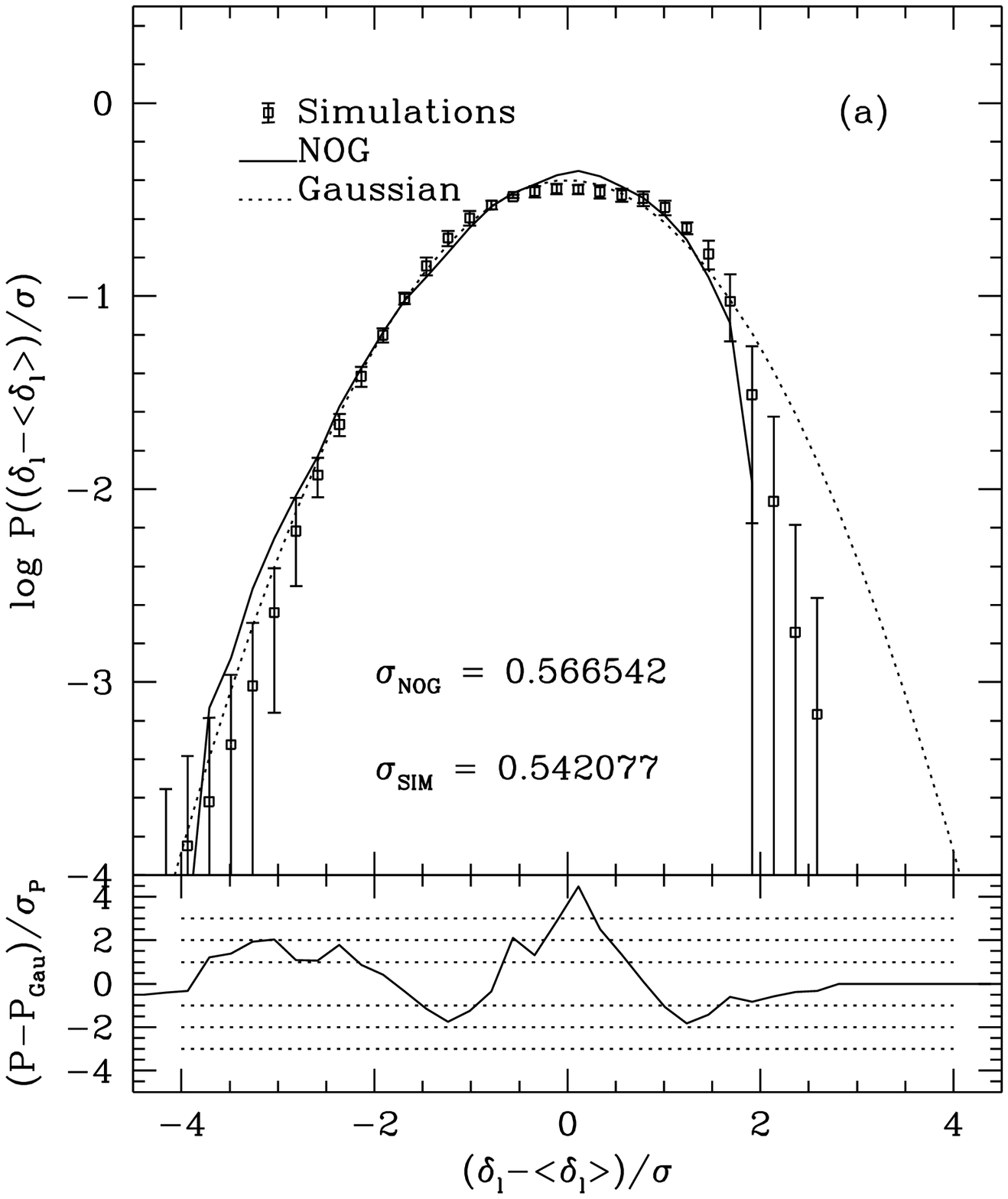,width=9cm}
\psfig{figure=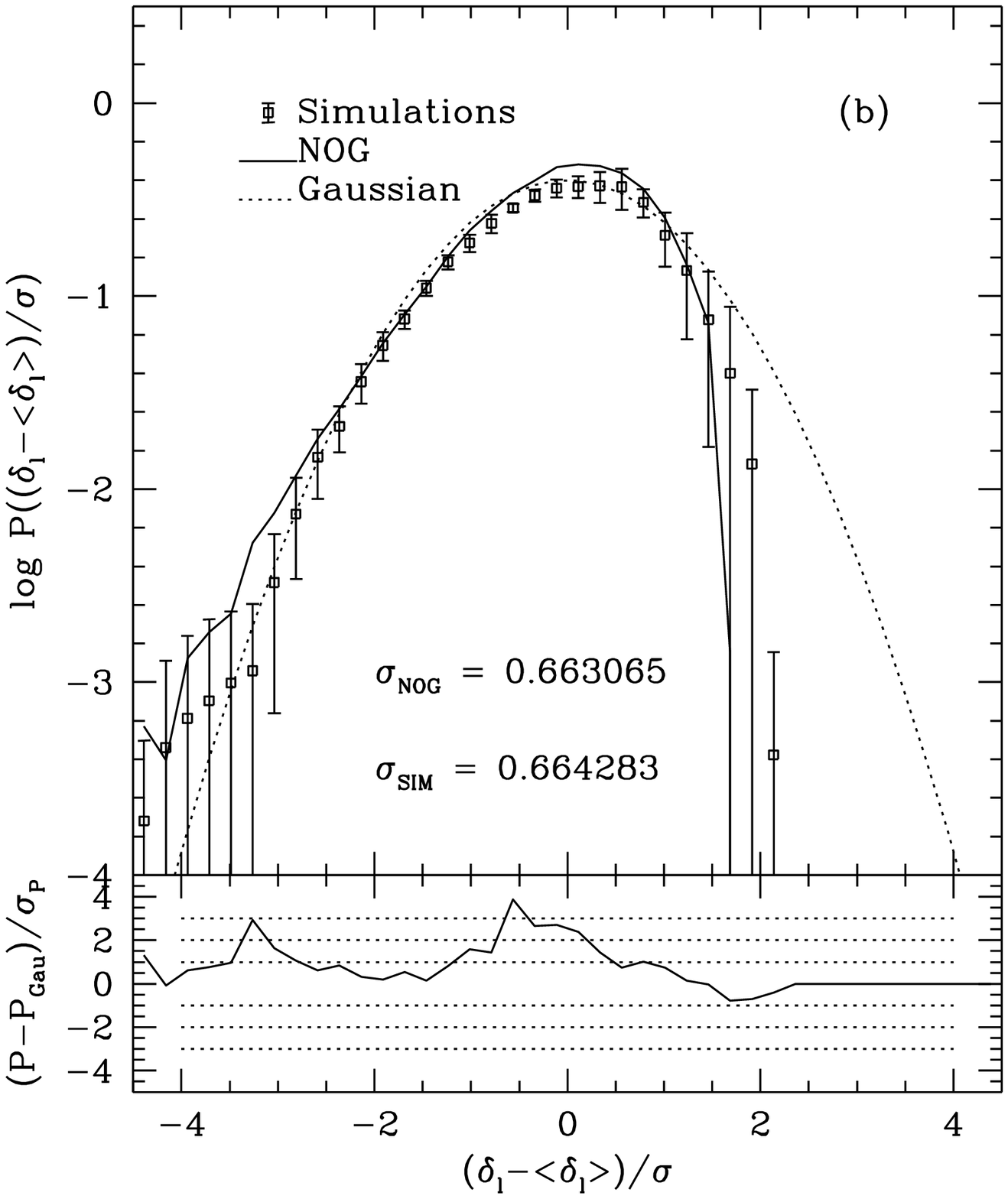,width=9cm}
}
\centerline{
\psfig{figure=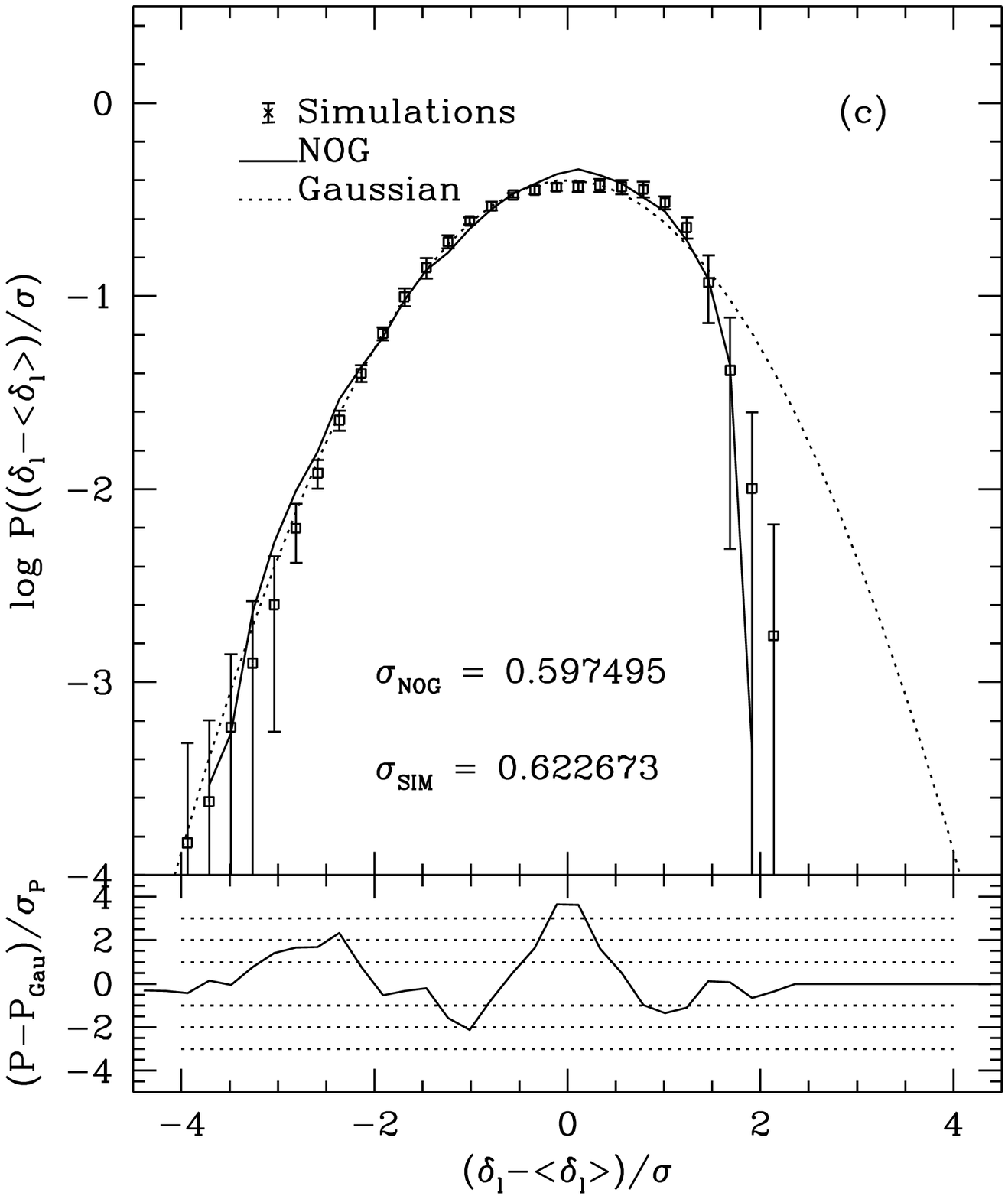,width=9cm}
\psfig{figure=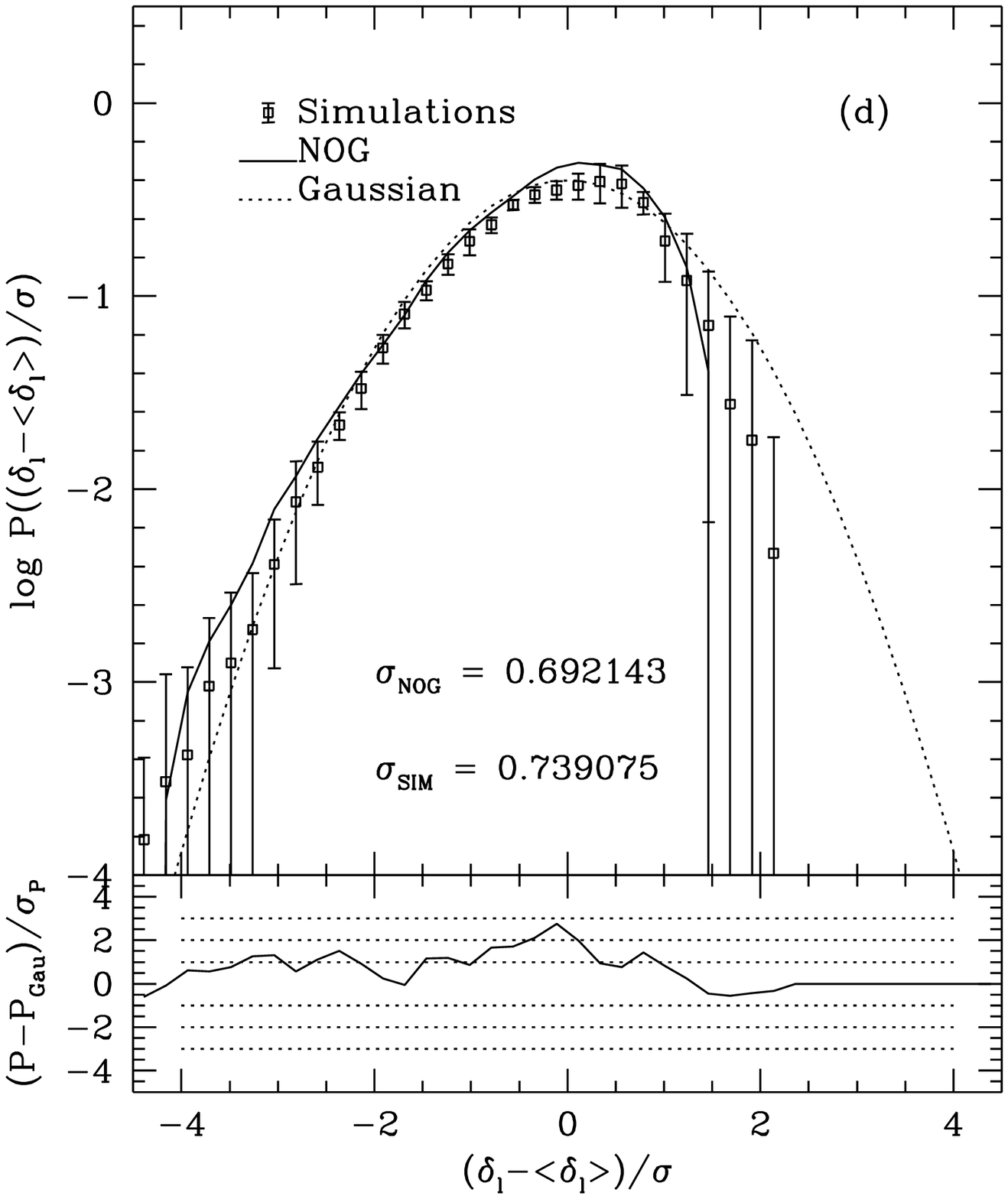,width=9cm}
}
\caption{Reconstructed PDFs. The upper panels (a,b) show the
reconstructed PDFs for the NOG catalogues filled assuming $b_{\rm
rel}=1.2 $ (continuous line) compared with the average PDFs computed
from the biased NOG catalogues filled with unbiased PSCz simulated
catalogues (points with errorbars). The left panel (a) shows the case
for $d_{\rm max}=60$ \hmpc, the right panel (b) shows the case for
$d_{\rm max}=30$ \hmpc. The lower panels (c,d) show the reconstructed
PDFs for the NOG catalogues filled assuming $b_{\rm rel}=1$ (continuous
line) compared with the average PDFs computed from the unbiased NOG
catalogues filled with unbiased PSCz simulated catalogues (points with
errorbars). The left panel (c) shows the case for $d_{\rm max}=60$
\hmpc, the right panel (d) shows the case for $d_{\rm max}=30$
\hmpc. Dotted lines show a Gaussian with zero mean and unit variance.
In each panel the residuals of the reconstructed PDFs with respect to
that of simulations, rescaled to the variance among the simulations,
are shown below.}
\label{fig:pdf}
\end{figure*}

Figure~\ref{fig:pdf} shows the reconstructed PDFs for the two NOG
catalogues (filled assuming either $b_{\rm rel}=1$ or 1.2) for $d_{\rm
max}=30$ or 60 \hmpc; at this stage the flattening of the peaks and
the restoration of power lost by smoothing have not been applied yet.
The curves are compared with the average PDFs computed from the
simulated catalogues.  Consistently with Monaco et al. (2000), no
distortion of the PDF is detected beyond that induced by ZTRACE.
There is a small discrepancy at $\delta\sim 0$, but it is valid only
for one or at most two points, and does not resemble any of the
patterns induced by realistic bias schemes (see Monaco et al. 2000);
we do not consider this feature as significant.

\begin{figure*}
\centerline{
\psfig{figure=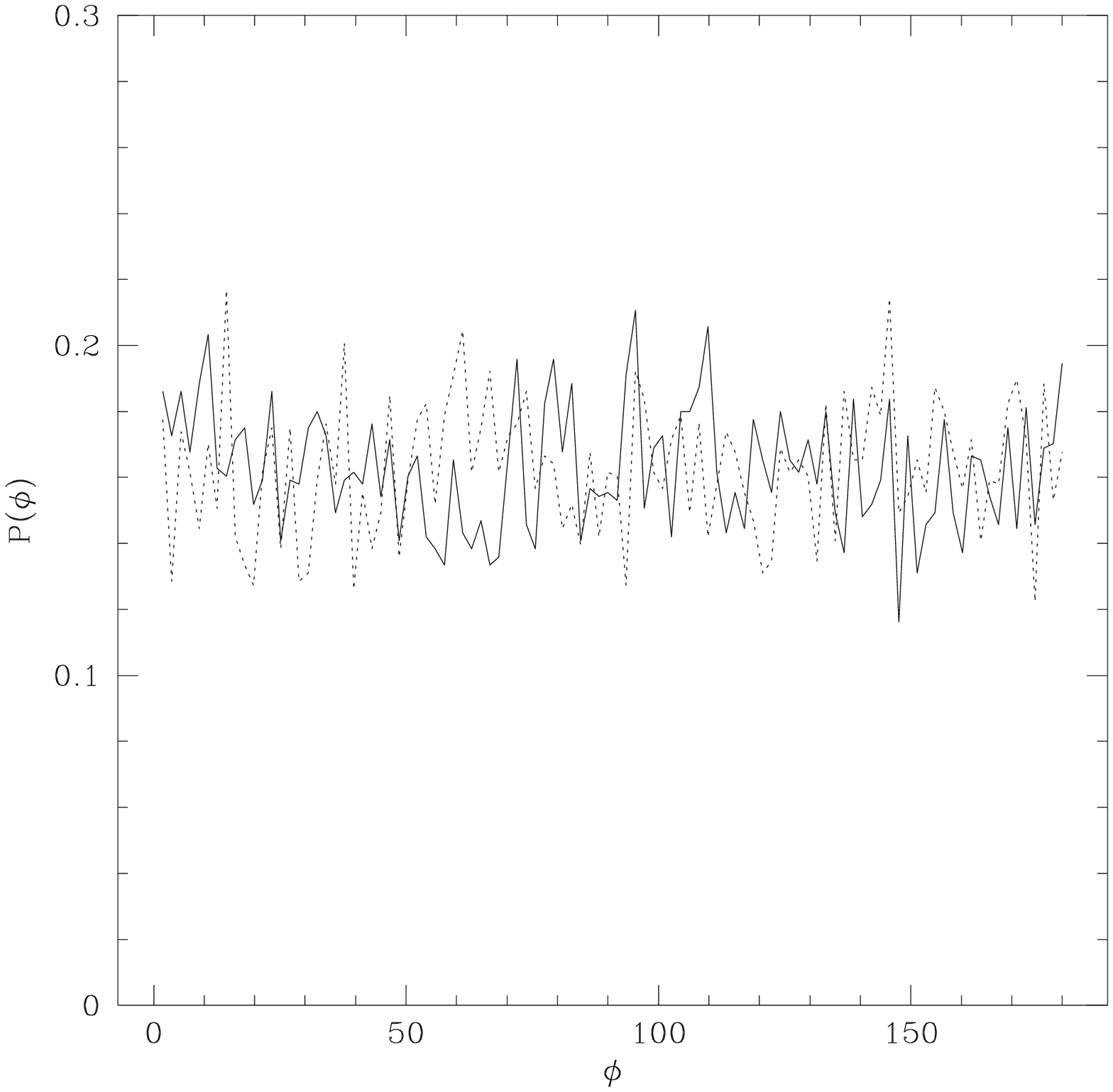,width=9cm}
\psfig{figure=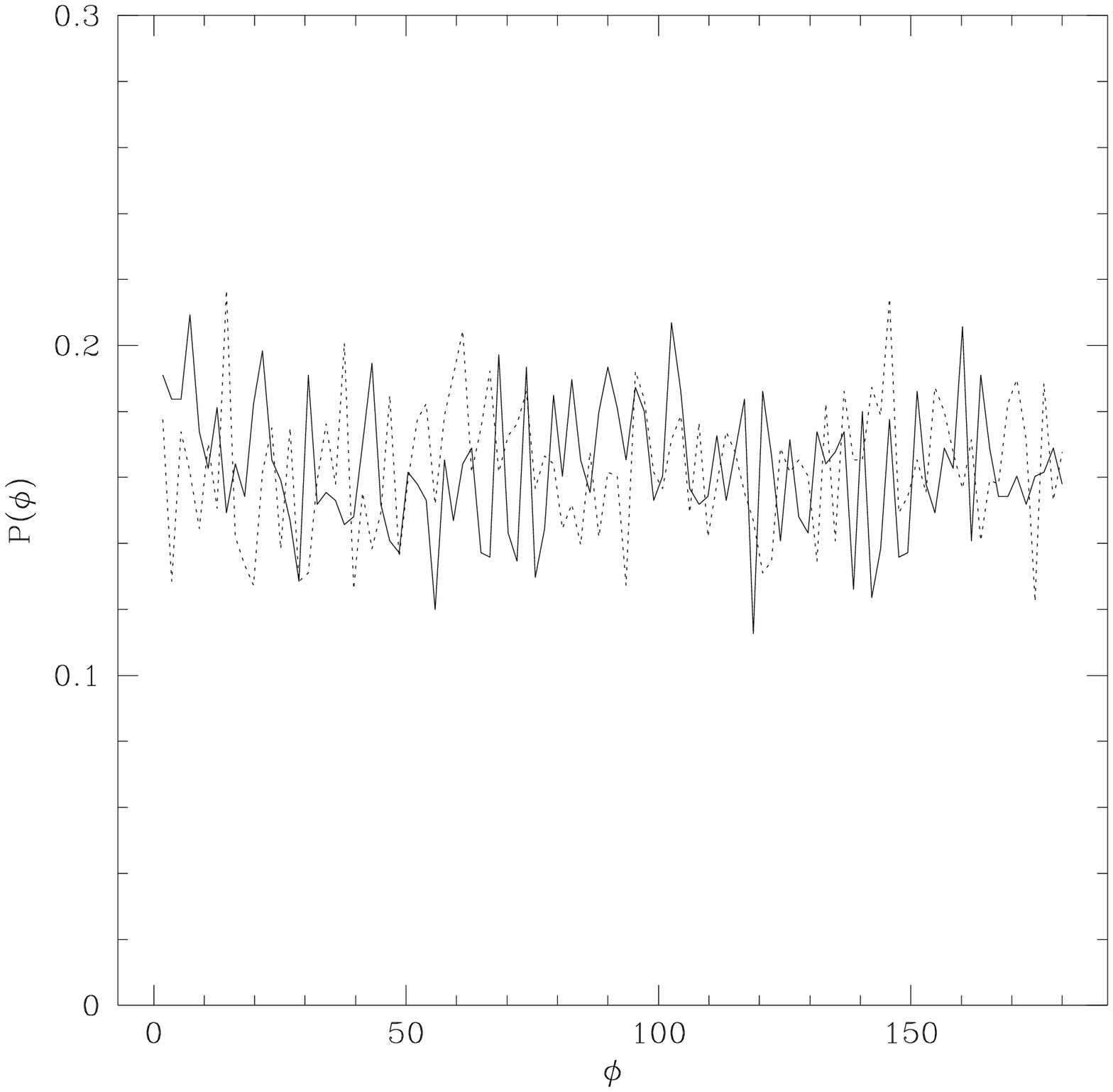,width=9cm}
}
\caption{Distribution of the phases of reconstructed initial
conditions for the two completions of NOG (continuous lines), compared
with the initial condition from one simulated catalogue (dotted
lines). Fields are padded to zero beyond $d_{\rm max}=60$ \hmpc. Right
panel refers to the case $b_{\rm rel}=1$, left panel refers to the
case $b_{\rm rel}=1.2$.}
\label{fig:fasinog}
\end{figure*}

Finally, figure~\ref{fig:fasinog} shows the phase distribution of the
two sets of initial conditions padded to zero beyond $d_{\rm max}=60$
\hmpc, for $k<k_{\rm lim}$, where $k_{\rm lim}$ is taken from the
corresponding case of table~\ref{table:tabellona60}.  For sake of
comparison we show also the phase distribution of the initial
conditions padded beyond $d_{\rm max}$ for one of the simulated
catalogue centres. It is apparent that the weak difference from
flatness of the reconstructed phases is consistent with it being an
effect of padding (see also figure~\ref{fig:fasidist}).  A similar
result holds for the PSCz reconstruction.

\subsection{Simulations of the local Universe}

The three reconstructions of the initial conditions of our local
Universe have been fixed for flattened peaks, as described in section
3.3, and for smoothing, as described in section 3.6. They have then
been FFT-transformed and truncated at $k_{\rm lim}$.  Finally, the
modes at $k>k_{\rm lim}$ have been extracted assuming a $\Lambda$CDM
power spectrum with the parameters given in section 3.1 and random
phases.  

The so obtained density field can be used to generate displacements
for a set of particles, so as to produce initial conditions for a
constrained N-body simulation of our local Universe.  A more refined
way to do it relies on the Hoffmann \& Ribak (1991, 1992) approach to
generate constrained realizations.  This would produce a Gaussian
density field that is not oversmoothed at large distances.  However,
in a multi-grid simulation the density field at large distance would
anyway be sampled by rather massive particles, thus removing the
intermediate scale recovered by the Hoffmann-Riback method.

A comparison of a full-blown simulation with the properties of our
local Universe, like the analysis of Narayanan et al. (2001), would
require a careful assessment of the bias relation between dark matter
and galaxies, and is considered beyond the scope of this paper.
However, to perform a ``quick and dirty'' check of the goodness of the
reconstruction, we follow the evolution of 30 different realizations
of the initial field (ten for each different reconstructed catalogue)
using the PINOCCHIO tool, recently proposed by Monaco et al.
(2002)\footnote{http://www.daut.univ.trieste.it/pinocchio/}.  This
gives a very fast and accurate approximation of the final distribution
of dark matter halos in terms of masses, positions, velocities and
merger histories, and has the advantage of running in a very short
time compared to a standard N-body code.  We are interested in the
reconstruction of the large-scale structure, so high resolution is not
required for this test.  For this reason we run PINOCCHIO on 128$^3$
grids, for a mass resolution of $\sim 10^{12}\ M_\odot$ ($h=0.7$).  In
this case, the smallest reliable halo reconstructed by PINOCCHIO (as
well as by an equivalent simulation) is as massive as $\sim 3-5\times
10^{13}\ M_\odot$; this mass scale is still dominated by the
small-scale power which is randomly added to the ZTRACE
reconstruction.

\begin{figure}
\centerline{
\psfig{figure=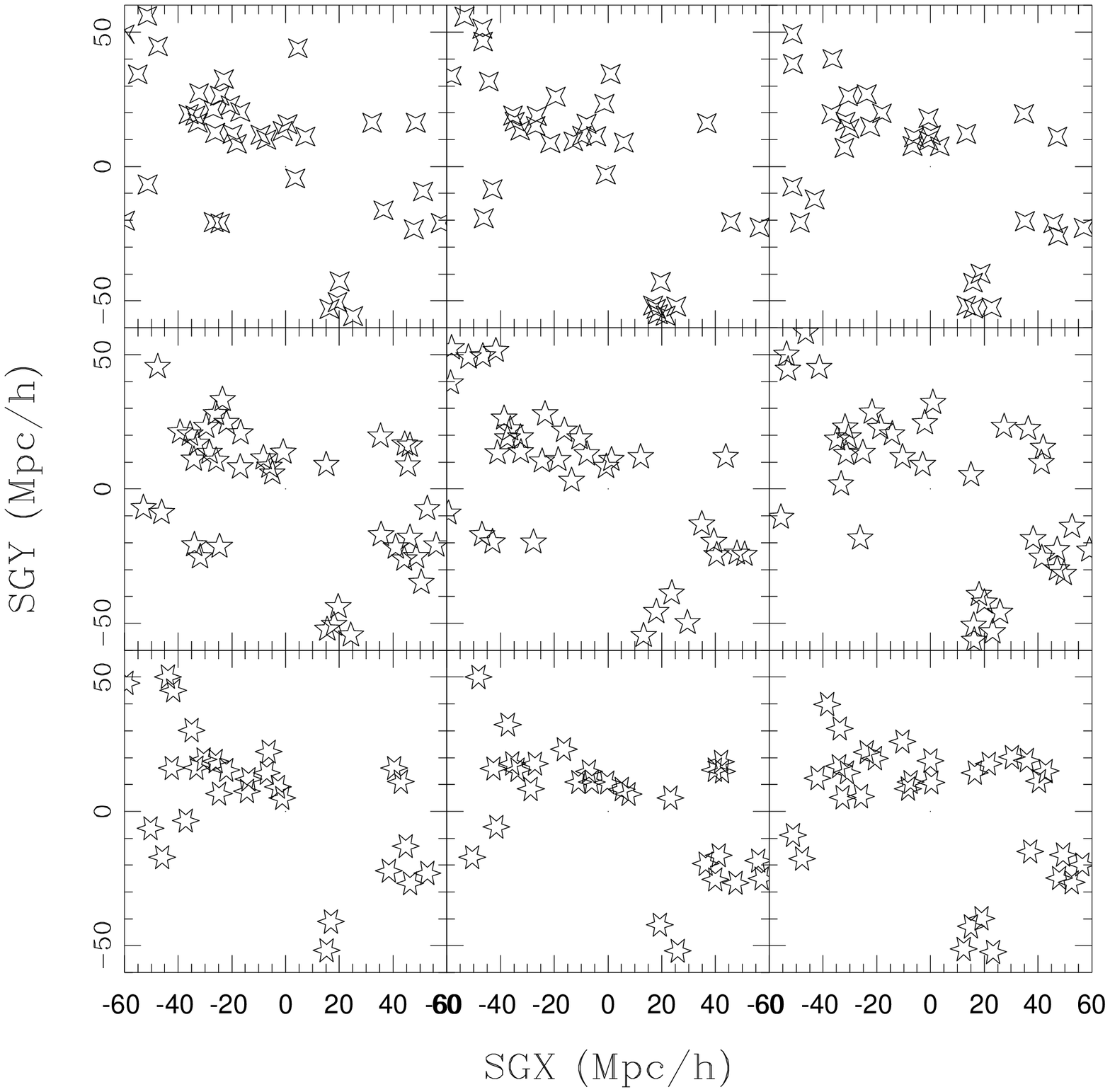,width=9cm}
}
\caption{Reconstruction of structures along (left panels) and
perpendicular (right panels) to the supergalactic plane. Upper panels
show 3 realizations based on NOG with $b_{\rm rel}=1.2$; mid panels
show 3 realizations based on NOG with $b_{\rm rel}=1$; lower panels
show 3 realizations based on PSCz. Stars indicate halos more massive
than $4\times 10^{13}\ M_\odot$.}
\label{fig:sgp}
\end{figure}

\begin{figure}
\centerline{
\psfig{figure=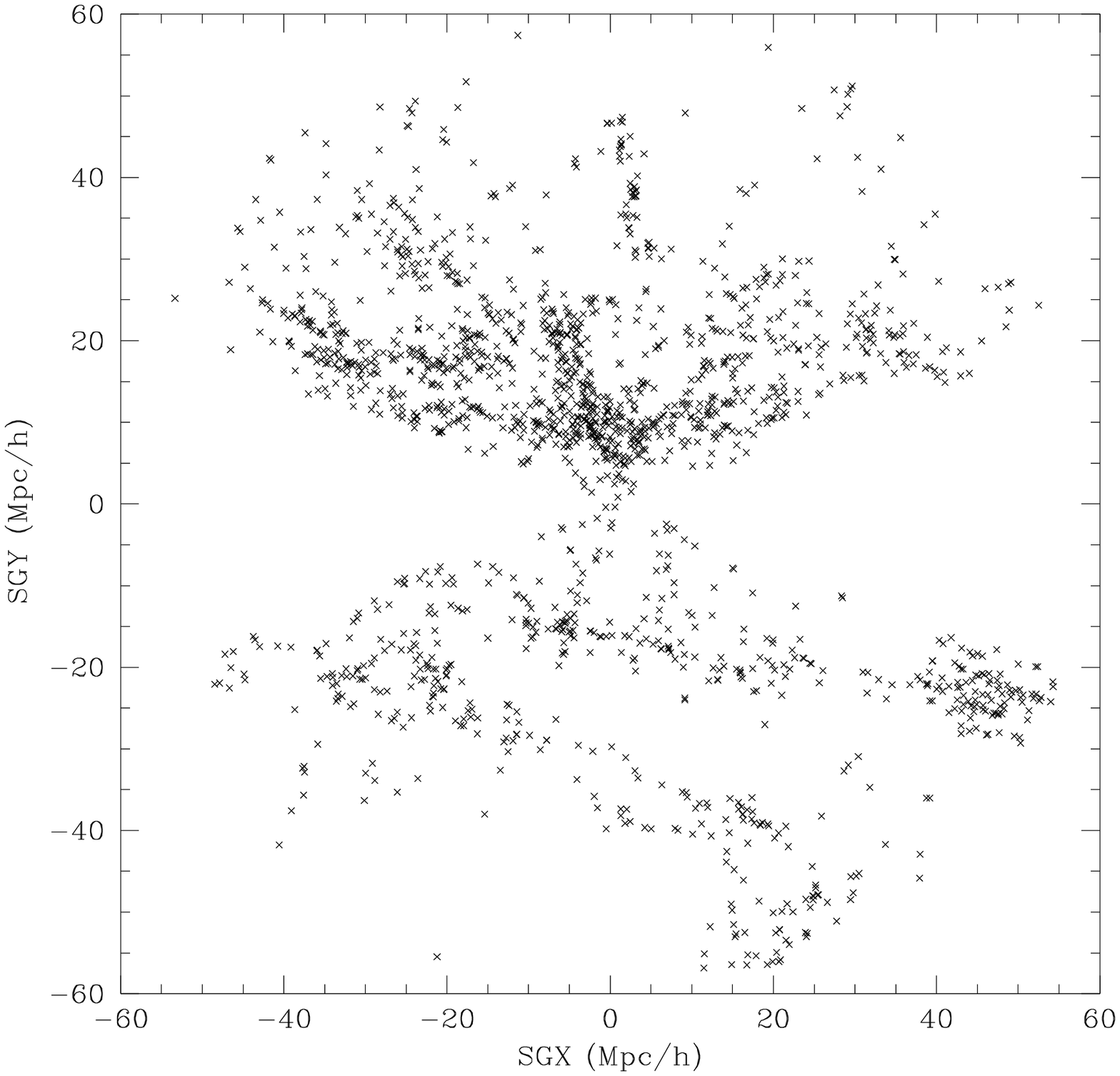,width=9cm}
}
\caption{Observed structures along the supergalactic plane. Panel
show the location of NOG galaxies.}
\label{fig:nogpscz}
\end{figure}

Figure~\ref{fig:sgp} shows the reconstruction of the NOG volume along
the super-galactic plane. Upper panels show 3 realizations based on
the NOG catalogue with $b_{rel}=1.2$; mid panels show the same for
$b_{rel}=1.2$; lower panels show the same for the PSCz
catalogue. Stars indicate the positions of halos reconstructed by
PINOCCHIO, with mass grater than $8 \times 10^{13}$. For sake of
comparison we report in figure~\ref{fig:nogpscz} the location of NOG
galaxies. The main structures observed in the supergalactic plane are
correctly reconstructed.  In particular, we find that the locations of
the main neighbour clusters, like Virgo, Perseus, Pisces, Centaurus
and Hydra, are recovered in most if not all cases within a few \hmpc.
We have also checked that the sheet-like overdensity corresponding to
the supergalactic plane is well reproduced.  We find that, as
expected, the NOG reconstruction is better than PSCz at high galactic
latitude, while it is worse in the optical ZOA.

We conclude that the two catalogues provide comparable reconstructions
of the local Universe.

\section{Summary and Conclusions} 

In this paper we have applied the ZTRACE algorithm (Monaco \&
Efstathiou, 1999) to the NOG (Giuricin et al., 2000) and PSCz
(Saunders et al. 2000) catalogues to reconstruct the initial condition
of our Local Universe.  The use of two different catalogues allows us
to test and improve the reconstruction; in particular, the optical
catalogue presents the advantages of denser sampling and better
relation between optical light and galaxy mass, while the infra-red
catalogue benefits from more homogeneous selections and larger sky
coverage.

We have studied the relative bias between these catalogues, finding a
good linear relation between the density fields traced by them,
corresponding to a relative bias of $b_{\rm rel}=1.1\pm 0.1$, in
agreement with the expectation based on the ratio of the $\sigma_8$
parameters for optical and infra-red galaxies.  The scatter around
this relation is found to be larger than that implied by shot noise.
This may be a sign of stochastic bias; in this hypothesis we obtain a
value $\sim 0.25$ for the parameter $\sigma_b$ defined by Dekel \&
Lahav (1999).

We have quantified the accuracy of the reconstruction through
extensive testing with simulated galaxy catalogues.  The analysis has
been performed in the Fourier space.  The initial conditions
reconstructed by ZTRACE reproduce on large scales both the
fluctuations of the modules of the Fourier modes around the mean (the
power spectrum) and their phases.  The accuracy of the reconstruction
drops at a scale corresponding to $k_{\rm lim}\simeq 0.4\ h/Mpc$,
beyond which all information is lost.  This $k_{\rm lim}$ depends only
weakly on catalogue, bias scheme, filling of the ZOA and smoothing
scale.  In particular, while for a smoothing radius of $\sim 5-6$
\hmpc\ the loss of information is determined mainly by smoothing, the
reconstruction at smaller scales is limited by non-linearity.
Moreover, the denser sampling of the NOG catalogue gives only a
modest improvement to the reconstruction, which is anyway compensated
by the errors induced by the procedure to fill the ZOA.

Consistently with the PSCz case (Monaco et al. 2000), the initial
conditions of the NOG catalogue are found to be consistent with the
Gaussian, with no convincing distortion beyond the flattening of the
high peaks induced by ZTRACE.  We have also checked the distribution
of the phases, which is found to be consistent with a uniform one both
for PSCz and NOG.

The reconstructed initial conditions have been corrected for smoothing
and peak flattening, using corrections calibrated on the simulated
catalogues.  From these we have generated sets of initial conditions
suitable for N-body simulations by randomly generating the modes
beyond $k_{\rm lim}$ with a power spectrum given by the cosmology in
section 3.1.  We have used the PINOCCHIO tool to run three sets of 10
low-resolution simulations, one set from the PSCz and two sets for the
NOG (with the ZOA filled assuming two different values for the
relative bias between NOG and PSCz, namely $b_{\rm rel}=1$ and 1.2).
In all cases the large-scale structure observed in our local Universe
is correctly reproduced.  The locations of the main neighbouring
clusters are also reproduced within a few \hmpc.  We notice that the
NOG-based realizations are more accurate than PSCz at high galactic
latitude, while the opposite is true when the optical ZOA is
approached.

With respect to previous works, we improve in many regards.  We use
different galaxy catalogues, NOG and PSCz, in place of the smaller 1.2
Jy one (Fisher et al. 1995), used by Kolatt et al. (1996) and Mathis
et al. (2002); Klypin et al. (2001) used the MARK III catalogue of
peculiar velocities (Willick et al., 1997), which allows a
reconstruction based on linear theory, but at the cost of a much
noisier set of data.  Moreover, we use ZTRACE to self-consistently
obtain the initial density field from the redshift-space catalogues,
and then correct for peak flattening and smoothing with procedures
carefully calibrated on simulations.  

\section*{Acknowledgments}
This work would not have been realized without the support of Giuliano
Giuricin.  We thank George Efstathiou, Christian Marinoni and Saleem
Zaroubi for helpful comments and discussions, and Volker Springel for
making the GADGET code publicly available.  The numerical simulation
has been run on the IBM SP3 machine at the Centro di Calcolo of the
University of Trieste.

{}

\appendix
\end{document}